\pdfoutput=1
\documentclass[preprint,review,12pt]{elsarticle}

\usepackage{amssymb}
\usepackage{lineno}

\journal{Nuclear Instruments and Methods A}

\begin{document}

\begin{frontmatter}

%% Title, authors and addresses

%% use the tnoteref command within \title for footnotes;
%% use the tnotetext command for the associated footnote;
%% use the fnref command within \author or \address for footnotes;
%% use the fntext command for the associated footnote;
%% use the corref command within \author for corresponding author footnotes;
%% use the cortext command for the associated footnote;
%% use the ead command for the email address,
%% and the form \ead[url] for the home page:
%%
%% \title{Title\tnoteref{label1}}
%% \tnotetext[label1]{}
%% \author{Name\corref{cor1}\fnref{label2}}
%% \ead{email address}
%% \ead[url]{home page}
%% \fntext[label2]{}
%% \cortext[cor1]{}
%% \address{Address\fnref{label3}}
%% \fntext[label3]{}

\title{Near-Intrinsic Energy Resolution for 30 to 662 keV Gamma Rays in a High Pressure Xenon Electroluminescent TPC}

%% use optional labels to link authors explicitly to addresses:
%% \author[label1,label2]{<author name>}
%% \address[label1]{<address>}
%% \address[label2]{<address>}

\author[ific]{V.~\'Alvarez}
\author[coimbra]{F.I.G.M.~Borges}
\author[ific]{S.~C\'arcel}
\author[zaragoza]{J.~Castel}
\author[zaragoza]{S.~Cebri\'an}
\author[ific]{A.~Cervera}
\author[coimbra]{C.A.N.~Conde}
\author[zaragoza]{T.~Dafni}
\author[coimbra]{T.H.V.T.~Dias}
\author[ific]{J.~D\'iaz}
\author[lbnl]{M.~Egorov}
\author[i3m]{R.~Esteve}
\author[jinr]{P.~Evtoukhovitch}
\author[coimbra]{L.M.P.~Fernandes}
\author[ific]{P.~Ferrario}
\author[i3N]{A.L.~Ferreira}
\author[coimbra]{E.D.C.~Freitas}
\author[lbnl]{V.M.~Gehman}
\author[ific]{A.~Gil}
\author[lbnl]{A.~Goldschmidt\corref{cor1}}\ead{agoldschmidt@lbl.gov}
\author[zaragoza]{H.~G\'omez}
\author[ific]{J.J.~G\'omez-Cadenas}
\author[zaragoza]{D.~Gonz\'alez-D\'iaz}
\author[bogota]{R.M.~Guti\'errez}
\author[iowa]{J.~Hauptman}
\author[santiago]{J.A.~Hernando Morata}
\author[zaragoza]{D.C.~Herrera}
\author[zaragoza]{F.J.~Iguaz}
\author[zaragoza]{I.G.~Irastorza}
\author[bogota]{M.A.~Jinete}
\author[autonoma]{L.~Labarga}
\author[ific]{I.~Liubarsky}
\author[coimbra]{J.A.M.~Lopes}
\author[ific]{D.~Lorca}
\author[bogota]{M.~Losada}
\author[zaragoza]{G.~Luz\'on}
\author[i3m]{A.~Mar\'i}
\author[ific]{J.~Mart\'in-Albo}
\author[ific]{A.~Mart\'inez}
\author[lbnl]{T.~Miller}
\author[jinr]{A.~Moiseenko}
\author[ific]{F.~Monrabal}
\author[coimbra]{C.M.B.~Monteiro}
\author[i3m]{F.J.~Mora}
\author[aveiro]{L.M. Moutinho}
\author[ific]{J.~Mu\~noz Vidal}
\author[coimbra]{H.~Natal da Luz}
\author[bogota]{G.~Navarro}
\author[ific]{M.~Nebot}
\author[lbnl]{D.~Nygren}
\author[lbnl]{C.A.B.~Oliveira}
\author[politecnica]{R.~Palma}
\author[ift]{J.~P\'erez}
\author[politecnica]{J.L.~P\'erez Aparicio}
\author[lbnl]{J.~Renner}
\author[girona]{L.~Ripoll}
\author[zaragoza]{A.~Rodr\'iguez}
\author[ific]{J.~Rodr\'iguez}
\author[coimbra]{F.P.~Santos}
\author[coimbra]{J.M.F.~dos Santos}
\author[zaragoza]{L.~Segui}
\author[ific]{L.~Serra}
\author[lbnl]{D.~Shuman}
\author[ific]{A. Sim\'on}
\author[tamu]{C.~Sofka}
\author[ific]{M.~Sorel}
\author[i3m]{J.F.~Toledo}
\author[zaragoza]{A.~Tom\'as}
\author[girona]{J.~Torrent}
\author[jinr]{Z.~Tsamalaidze}
\author[santiago]{D.~V\'azquez}
\author[aveiro]{J.F.C.A.~Veloso}
\author[zaragoza]{J.A.~Villar}
\author[tamu]{R.C.~Webb}
\author[tamu]{J.T~White}
\author[ific]{N.~Yahlali}
\address[ific]{Instituto de F\'isica Corpuscular (IFIC), CSIC \& Universitat de Val\`encia,
Calle Catedr\'atico Jos\'e Beltr\'an, 2, 46980 Paterna, Valencia, Spain}
\address[coimbra]{Departamento de Fisica, Universidade de Coimbra, Rua Larga, 3004-516 Coimbra, Portugal}
\address[zaragoza]{Laboratorio de F\'isica Nuclear y Astropart\'iculas, Universidad de Zaragoza,
Calle Pedro Cerbuna 12, 50009 Zaragoza, Spain}
\address[i3m]{Instituto de Instrumentaci\'on para Imagen Molecular (I3M), Universitat Polit\`ecnica de Val\`encia,
Camino de Vera, s/n, Edificio 8B, 46022 Valencia, Spain}
\address[lbnl]{Lawrence Berkeley National Laboratory (LBNL),
1 Cyclotron Road, Berkeley, California 94720, USA}
\address[jinr]{Joint Institute for Nuclear Research (JINR),
Joliot-Curie 6, 141980 Dubna, Russia}
\address[aveiro]{Institute of Nanostructures, Nanomodelling and Nanofabrication (i3N), Universidade de Aveiro, Campus de Santiago, 3810-193 Aveiro, Portugal}
\address[bogota]{Centro de Investigaciones en Ciencias B\'asicas y Aplicadas, Universidad Antonio Nari\~no, Carretera 3 este No.\ 47A-15, Bogot\'a, Colombia}
\address[iowa]{Department of Physics and Astronomy, Iowa State University, 12 Physics Hall, Ames, Iowa 50011-3160, USA}
\address[santiago]{Instituto Gallego de F\'isica de Altas Energ\'ias (IGFAE), Univ.\ de Santiago de Compostela, Campus sur, R\'ua Xos\'e Mar\'ia Su\'arez N\'u\~nez, s/n, 15782 Santiago de Compostela, Spain}
\address[autonoma]{Departamento de F\'isica Te\'orica, Universidad Aut\'onoma de Madrid, Campus de Cantoblanco, 28049 Madrid, Spain}
\address[politecnica]{Dpto.\ de Mec\'anica de Medios Continuos y Teor\'ia de Estructuras, Univ.\ Polit\`ecnica de Val\`encia, Camino de Vera, s/n, 46071 Valencia, Spain}
\address[ift]{Instituto de F\'isica Te\'orica (IFT), UAM/CSIC, Campus de Cantoblanco, 28049 Madrid, Spain}
\address[girona]{Escola Polit\`ecnica Superior, Universitat de Girona, Av.~Montilivi, s/n, 17071 Girona, Spain}
\address[tamu]{Department of Physics and Astronomy, Texas A\&M University, College Station, Texas 77843-4242, USA}
\cortext[cor1]{Corresponding author}

\begin{abstract}
We present the design, data and results from the NEXT prototype for Double Beta and Dark Matter (NEXT-DBDM) detector, a 
high-pressure  gaseous natural xenon electroluminescent time projection chamber (TPC) that was built at the Lawrence Berkeley 
National Laboratory. It is a prototype of the planned NEXT-100 $^{136}$Xe neutrino-less double beta decay ($0\nu\beta\beta$) 
experiment with the main objectives of demonstrating near-intrinsic energy resolution at energies up to 662 keV and of 
optimizing the NEXT-100 detector design and operating parameters. Energy resolutions of $\sim$1\% FWHM for 662 keV gamma 
rays were obtained at 10 and 15 atm and $\sim$5\% FWHM for 30 keV fluorescence xenon X-rays. These results demonstrate that 0.5\% FWHM resolutions for the 2,459 keV hypothetical neutrino-less double beta decay peak are realizable. This energy resolution 
is a factor 7 to 20 better than that of the current leading $0\nu\beta\beta$ experiments using liquid xenon and 
thus represents a significant advancement. We present also first results from a track imaging system consisting of 64 silicon photo-multipliers recently installed in NEXT-DBDM that, along with the excellent energy resoution, demonstrates the key functionalities required for the 
NEXT-100 $0\nu\beta\beta$ search. 

\end{abstract}

\end{frontmatter}

%%
%% Start line numbering here if you want
%%
%%\linenumbers

%% main text
%% \section{}
%% \label{}

\section{Introduction}
\label{introduction}

Neutrino-less double beta decay ($0\nu\beta\beta$) is a postulated \cite{Majorana:1937} rare process in which a nucleus 
changes by two units of charge while emitting two electrons (or positrons) without the emission of neutrinos \cite{Cadenas:2012}.
Should this decay happen in nature the sum of the energies of the two electrons will be monoenergetic at exactly the 
Q-value of the nuclear decay ($Q_{\beta\beta}$, equal to the mass difference between the parent and daughter nuclei).  A 
precise energy measurement can therefore greatly aid in the identification of the $0\nu\beta\beta$ process in the presence 
of other more common processes that produce either continuous energy deposition spectra or peaks at well known and well 
separated energies. The occurrence of $0\nu\beta\beta$  would imply  that neutrinos are their own 
anti-particle \cite{Racah:1937, Schechter:1982}, or Majorana particles. Should neutrinos prove to be Majorana particles 
the observed prevalence of matter over anti-matter in our universe could be explained through the Leptogenesis 
mechanism (see e.g. \cite{Buchmuller:2005, Pilaftsis:2009}). 

The $^{136}$Xe isotope is one of the candidate nuclei in which $0\nu\beta\beta$, in this case $^{136}$Xe $\rightarrow$ 
$^{136}$Ba + $e^-$ + $e^-$, could be detectable because the single beta decay which would otherwise dominate the experimental 
count rate is energetically forbidden. Two currently running experiments are searching for this rare process in $^{136}$Xe. 
EXO \cite{EXO:2012} uses 200 kg of the enriched isotope in a cryogenic liquid xenon TPC and KamLAND-Zen \cite{KamLANDZen:2012} 
uses 330 kg of the isotope dissolved in 13 tons of organic scintillator. 
The energy resolutions for these two experiments are 3-4\% and 10\% FWHM respectively at the 2.459 MeV \cite{McCowan:2010} 
$Q_{\beta\beta}$ of the $^{136}$Xe decay. It is well known, on the other hand, that xenon in gaseous phase can offer 
significantly better energy resolution due to its small Fano factor \cite{Fano:1947} F = 0.14 (a measure of the 
level of fluctuations in the number of ionization electrons). For a xenon gas pressure of less than 57 atm at room temperature (density of 0.55 g/cm$^3$) the intrinsic energy resolution is expected to be \cite{Bolotnikov:1997} about  0.3\% FWHM near the $Q_{\beta\beta}$. It is thus clear that a xenon detector at 
moderately high pressure would represent a significant advantage for the search of the $0\nu\beta\beta$ spectrum peak as long 
as its implementation can preserve a near-intrinsic energy resolution.

NEXT-100 is an experiment \cite{NEXTTDR:2012} being constructed to search for $0\nu\beta\beta$ using 100-150 kg of 
$^{136}$Xe in a 10-15 atm TPC at the Canfranc Underground Laboratory (LSC) under the Pyrenees mountains in Spain. 
In the TPC, conceptually developed in Ref. \cite{Nygren:2009}, electrons liberated through ionization by the passage of energetic charged particles (such as the two electrons from the $0\nu\beta\beta$ decay) drift under the presence of a weak electric field towards a thin region with a high electric 
field. The {\it E/P} (electric field divided by pressure) in this high field region is such that electrons acquire enough energy 
to excite xenon atoms, but not enough to ionize them. Most of the excitation energy is ultimately released in the form of 
ultra-violet (VUV) photons of wavelengths near 172 nm and constitutes the electroluminescence (EL) \cite{Noblegasdetectors:2006} 
signal. 
For each ionization electron, thousands of EL photons can be produced in the EL amplification region, with extremely low fluctuations in the number of EL photons produced. An array of photomultiplier tubes (PMTs) then detects a fraction of the VUV photons to render a measurement of the total energy
released in the gas with a statistical precision near the Fano limit.

The NEXT-100 TPC will provide, in addition to a very precise energy measurement, a 3-D image of the ionization tracks by 
means of a dense array of silicon photomultipliers \cite{NEXT_SiPMs:2012} (SiPMs or MPPCs) installed near the electroluminescence region. 
This topological information is useful to distinguish between events with 2 electrons emerging from a single point, such as 
in $0\nu\beta\beta$, from events with one or more electrons that result from single-site and multi-site interactions of gamma rays from
 natural radioactivity in the detector and surrounding materials.

In this paper we present the design, data and results from the NEXT-DBDM (NEXT prototype for Double Beta and Dark Matter) 
detector, a 1 kg natural xenon electroluminescent TPC that was built at the Lawrence Berkeley National Laboratory. It is a 
prototype of the NEXT-100 detector with the main objectives of demonstrating the near intrinsic energy resolution at energies 
up to 662 keV and of optimizing the NEXT-100 detector design, construction, and operating parameters.

\section{High Pressure Xenon Electroluminescent TPC}

In the past, various high pressure xenon detector designs have been created for studies directed towards the improvement of energy
resolution.  Studies of gridded ionization chambers in cylindrical form \cite{Bolotnikov:1997TNS} and with parallel plate geometries
\cite{Mahler:1998, Tepper:1995, Levin:1993, Bolotnikov:1997} were able to achieve energy resolutions at the level of 2-2.5\% FWHM at 662 keV.
Later studies investigated alternatives to shielding grids such as a hemispherical geometry \cite{Kessick:2002}, using a virtual Frisch grid
\cite{Bolotnikov:2004}, pulse correction using drift time determined by observing primary scintillation \cite{Lacy:2004}, and variations 
\cite{Bolotnikov:2004, Kiff:2005} of the coplanar anode approach introduced in \cite{Luke:1994}.  These studies produced energy resolutions of 
2-6\% FWHM at 662 keV.  It has been shown in \cite{Bolozdynya:1997} that good energy resolution (2.7\% FWHM at 122 keV) is achievable using 
electroluminescent readout (see \cite{Nygren:2009} for a more thorough historical account of electroluminescence in detectors), and here we 
extend the usage of this technique to higher energies.

The basic building blocks of the NEXT-DBDM xenon electroluminescent TPC are: a stainless steel pressure vessel, a gas system 
that recirculates and purifies the xenon at 10-15 atm, stainless steel wire meshes that establish high-voltage equipotential planes in 
the boundaries of the drift and the EL regions, field cages with hexagonal cross sections to establish uniform electric fields 
in those regions,  an hexagonal pattern array of 19 VUV sensitive PMTs inside the pressure vessel and an associated readout 
electronics and data acquisition (DAQ) system.

When ionizing radiation traverses the drift region of the TPC, xenon atoms are ionized or excited. Most of the excitation 
energy is promptly released as a fast scintillation pulse of 172 nm VUV photons that lasts 10-30 ns \cite{Noblegasdetectors:2006}. 
A fraction of these photons are detected in the PMT array, forming the S1 signal that provides the start time $t_0$ for the 
TPC. The ionization (or secondary) electrons, on the other hand, drift for a maximum distance of 8 cm at a velocity of 
$\sim$ 0.1 cm/$\mu$s towards the EL region. There, they accelerate and produce copious EL VUV photons. The same PMT array detects a fraction of these photons, forming the S2 signal.

\begin{figure}[!htb]
  \centering
\includegraphics[scale=0.10]{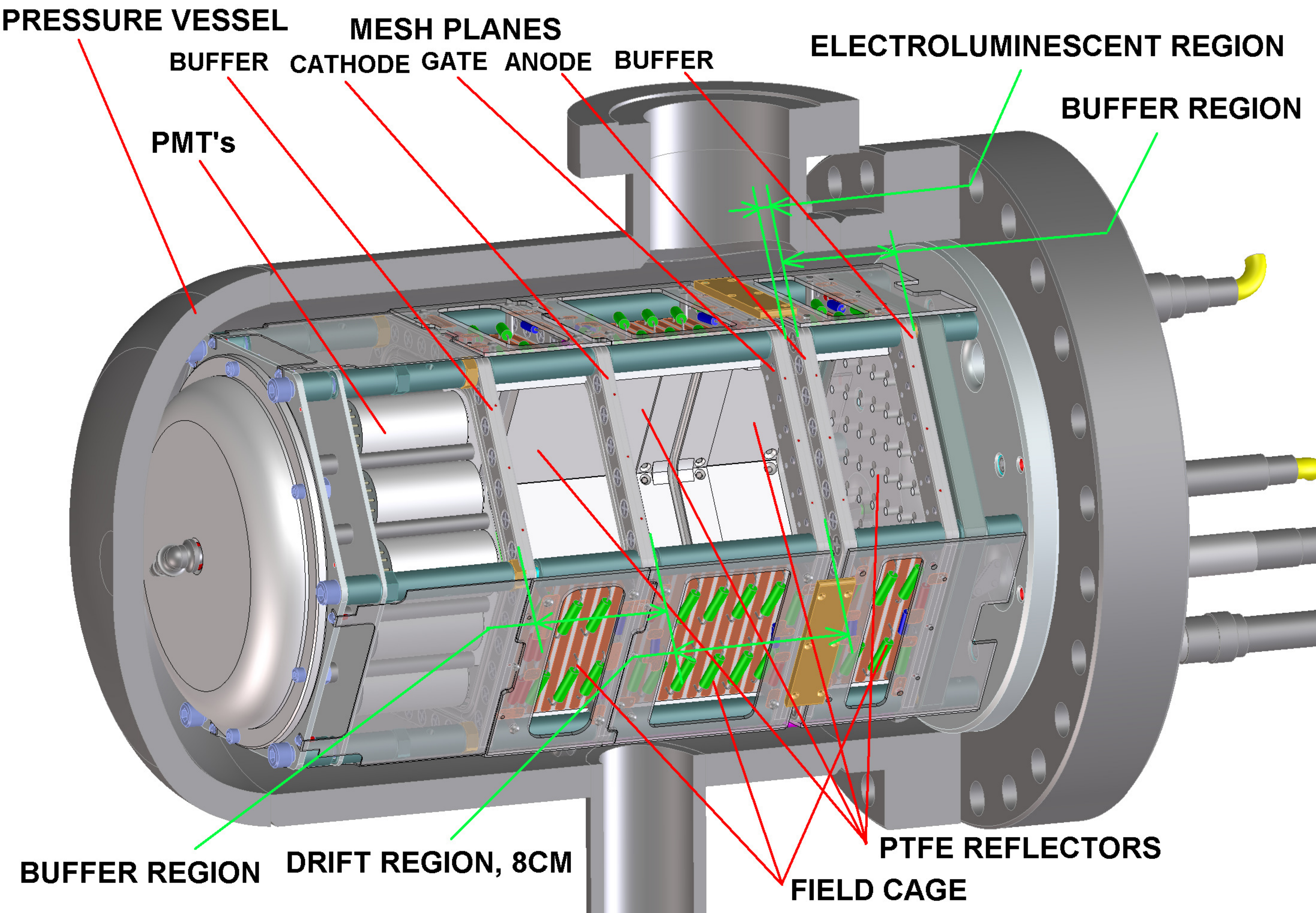}
  \caption{\label{tpc_configuration} \textbf{NEXT-DBDM electroluminescent TPC configuration:} An array of 19 photomultipliers (PMTs) measures S1 
  primary scintillation light from the 8 cm long drift region and S2 light produced in the 0.5 cm electroluminescence (EL) region 13.5 cm away 
  from the PMTs. Two 5 cm long buffer regions behind the EL anode mesh and between the PMTs and the cathode mesh grade the high voltages (up to 
  $\pm$17 kV) down to ground potential.}
\end{figure}

In the NEXT-DBDM detector the PMT array and the EL region, which are both hexagonal areas with 12.4 cm between opposite 
sides, are 13.5 cm away from each other (see Fig.\ref{tpc_configuration}). Thus point-like isotropic light produced in the 
EL region illuminates the PMT array with little PMT-to-PMT variation. This geometric configuration also makes the illumination pattern 
and the total light collection only very mildly dependent on the position of the light origin within the EL region. The 
diffuse reflectivity of the TPC walls increases this light collection uniformity further. As a result, the device provides 
good energy measurements with little dependence on the position of the charge depositions. On the other hand, without a light 
sensor array near the EL region precise tracking information is not available and only coarse average position can be obtained using the PMT array light pattern. In this report, only the tracking analysis in section \ref{sipm_section} was performed using data acquired with the tracking sensor array recently installed in the NEXT-DBDM TPC.  In a future publication we will report on the energy resolution achieved in a combined analysis using information from both the PMT and tracking arrays.

\section{Intrinsic Energy Resolution in the HPXe TPC}
The intrinsic energy resolution in a xenon gas detector that measures ionization is given by:

\begin{equation}
\delta E/E=2.35\sqrt{(FW_\mathrm{i}/E)} \mathrm{\quad(FWHM)}
\end{equation}

where $E$ is the energy released in the detector, $W_\mathrm{i}$ is the average energy required to liberate an electron and $F$ is the 
Fano factor that quantifies the fluctuations in the number of liberated electrons. 
$F$ and $W_\mathrm{i}$ are energy, drift field and pressure dependent due to electron-ion recombination and due to the energy dependence of the energy loss rate $dE/dx$ (see for example\cite{Bolotnikov:1997}, \cite{Dias:1992} and \cite{Dias:1997}). In this study we explored the 30-662 keV energy range at pressures between 10 and 17 atm, and drift fields in the  
0.3-2.0 kV/cm range. For the purpose of studying the energy resolution in the NEXT-DBDM detector we used 662 keV 
gamma rays from a $^{137}$Cs source and 29.1-34.5 keV xenon X-rays \cite{LBLTOI} that follow photoelectric interactions of the gamma rays. 
For $F$=0.14 \cite{Doke:2005} and $W_\mathrm{i}$=24.8 eV \cite{Platzman:1961,Nygren:2009} the intrinsic energy resolutions are approximately  0.53\% FWHM 
for 662 keV and 2.5\% FWHM for 30 keV.

\section{Experimental Aspects of the Energy Resolution in the Electroluminescent HPXe TPC}
\subsection{Statistical}
In the NEXT-DBDM detector the number of liberated electrons is not measured directly. Rather, those electrons are drifted towards the 
EL region and then accelerated to produce O(1000) VUV photons for each electron crossing the EL region of which O(10) are measured as photoelectrons in the PMT array. This gain and measurement sequence introduces fluctuations beyond the intrinsic Fano limit. In Ref. \cite{Borges:1999} a 
formalism was developed to calculate the energy resolution achievable in light of those additional fluctuations:

 \begin{equation}
\label{eq:resolution}
\delta E/E=2.35\sqrt{((F+G)W_\mathrm{i}/E)} \mathrm{\quad(FWHM)}
\end{equation}
 with 
\begin{equation}
\label{eq:G}
G=1/\eta+(1+\sigma^{2}_\mathrm{pd})/n_\mathrm{pe}
\end{equation}

where $\eta$ is the average number of VUV photons produced in the EL region per secondary electron (or optical gain), 
$n_\mathrm{pe}$ is the average number of photons detected (as photoelectrons) per secondary electron and $\sigma_\mathrm{pd}^{2}$ is the 
variance on the charge measured for single photoelectrons. \footnote{A thorough analysis of the statistics of energy resolution gives the formula 

 \begin{equation}
  \Biggl(\frac{\Delta E}{E}\Biggr) = 2.35 \sqrt{\frac{F}{n_\mathrm{ion}} + \frac{J_\mathrm{CP}-1}{n_\mathrm{EL}} + \frac{(\sigma_{q}/\bar{q})^{2} + 1}{n_\mathrm{det}}}\,\,\mathrm{\quad(FWHM)}.
 \end{equation}
 
 \noindent where $n_\mathrm{ion} = E/W_\mathrm{i}$ is the number of electron-ion pairs produced, $n_\mathrm{EL} = n_\mathrm{ion}\,\eta$ is the total number of EL photons, and
 $n_\mathrm{det} = \eta\,\varepsilon \,n_\mathrm{ion}$ is the total number of photoelectrons detected by the PMTs ($\varepsilon$ is the detection efficiency of EL photons).  $J_\mathrm{CP}$ is the Conde-Policarpo factor which characterizes the fluctuations in EL light production similar to how the Fano factor $F$ 
 characterizes the fluctuation in ionization production.  Equation \ref{eq:resolution} assumes $J_\mathrm{CP} = 2$. In most realistic scenarios, however, the second term under the square root can be neglected due to the low EL photon detection probability.
}

The electroluminescence optical gain in pure high pressure xenon is given approximately by \cite{Monteiro:2007, Nygren:2009}:
\begin{equation}
\label{eq:elgain}
\eta=140\left (\frac{\Delta V}{p\Delta z}-0.83\right )\cdot p \cdot \Delta z
\end{equation}

where $\Delta V$ is the voltage difference between the mesh planes that form the EL region in kilovolts, $p$ is the pressure 
in atmospheres and $\Delta z$ is the distance between the electrodes in centimeters. For instance, if $p$=10 atm, 
$\Delta V$=11.3 kV and $\Delta z$=0.5 cm the EL optical gain is $\eta$=1000.

The value of $n_\mathrm{pe}$ is the product of the EL optical gain $\eta$ times the collection efficiency (the probability of a VUV photon 
generated in the EL region reaching a PMT window) times the PMT quantum efficiency at the corresponding wavelength. 
For example for a 10\% collection efficiency, a 15\% PMT quantum efficiency at the 172 nm wavelength of the xenon electroluminescence and 
$\eta$=1000, $n_\mathrm{pe}$=15 and the expected energy resolution for 662 keV gamma rays in the absence of systematic effects is 
0.66\% and 3.11\% for 30 keV.
 
\subsection{Systematic}
Systematic effects that broaden the energy resolution can be grouped in two categories: position
and time dependencies. Position dependencies of the energy response arise, for example, from a non-uniform EL light collection efficiency,
a non-uniform EL gain,  secondary electron losses due to attachment on electronegative impurities during drift and secondary
electrons hitting the TPC walls due to diffusion or due to drift field non-uniformities. 
Time dependencies of the energy response arise, for example, from time-varying voltages, temperature (and its subsequent effect of gas viscosity, PMT response, etc), gas flow and purity, PMT response and gas density.    

In the NEXT-DBDM TPC systematic dependencies of the energy response as a function of the position along the drift ($z$-axis in our chosen coordinate system) are small but still non-negligible due to the high xenon gas purity achieved. In addition, the $z$ position of charge depositions within the drift region is very precisely measured. The opposite is true in the plane perpendicular to the drift direction ($x-y$): the light collection changes more rapidly as a function of distance from the center axis of the TPC and the $x-y$ position of charge depositions is poorly measured. For this reason, to study the energy resolution achievable with the xenon EL TPC, calibration gamma rays are introduced along the center $z$ axis through a narrow aperture collimator. Still, the actual charge depositions happen over an extended region in $x-y$ (and $z$) due to the length of the electron tracks and due to the multi-site depositions from Compton scatters and from xenon X-rays following photoelectric absorption.     
  
\section{Experimental Setup}

\subsection{Gas System}
The main functions of the gas system are to recirculate and purify the xenon at pressures up to 17 atm.
A magnetically driven seal-less and oil-less pressure rated (95 atm) pump manufactured by Pumpworks Inc. (model 2070)
recirculates the xenon at room temperature at 5-15 standard liter per minute (slpm). For the total system volume
of approximately 10 liter at 10 atm pressure the pump recirculates one full volume in about 10 minutes.
A pressure rated (18 atm) heated getter from Johnson Matthey (model PureGuard) removes O$_2$, H$_2$O, N$_2$
and many other impurities using a non-evaporable zirconium-based material. The getter operates at a constant 450 degrees Celsius
irreversibly removing the impurities through bulk diffusion.

\begin{figure}[!htb]
  \centering
 \includegraphics[scale=0.7]{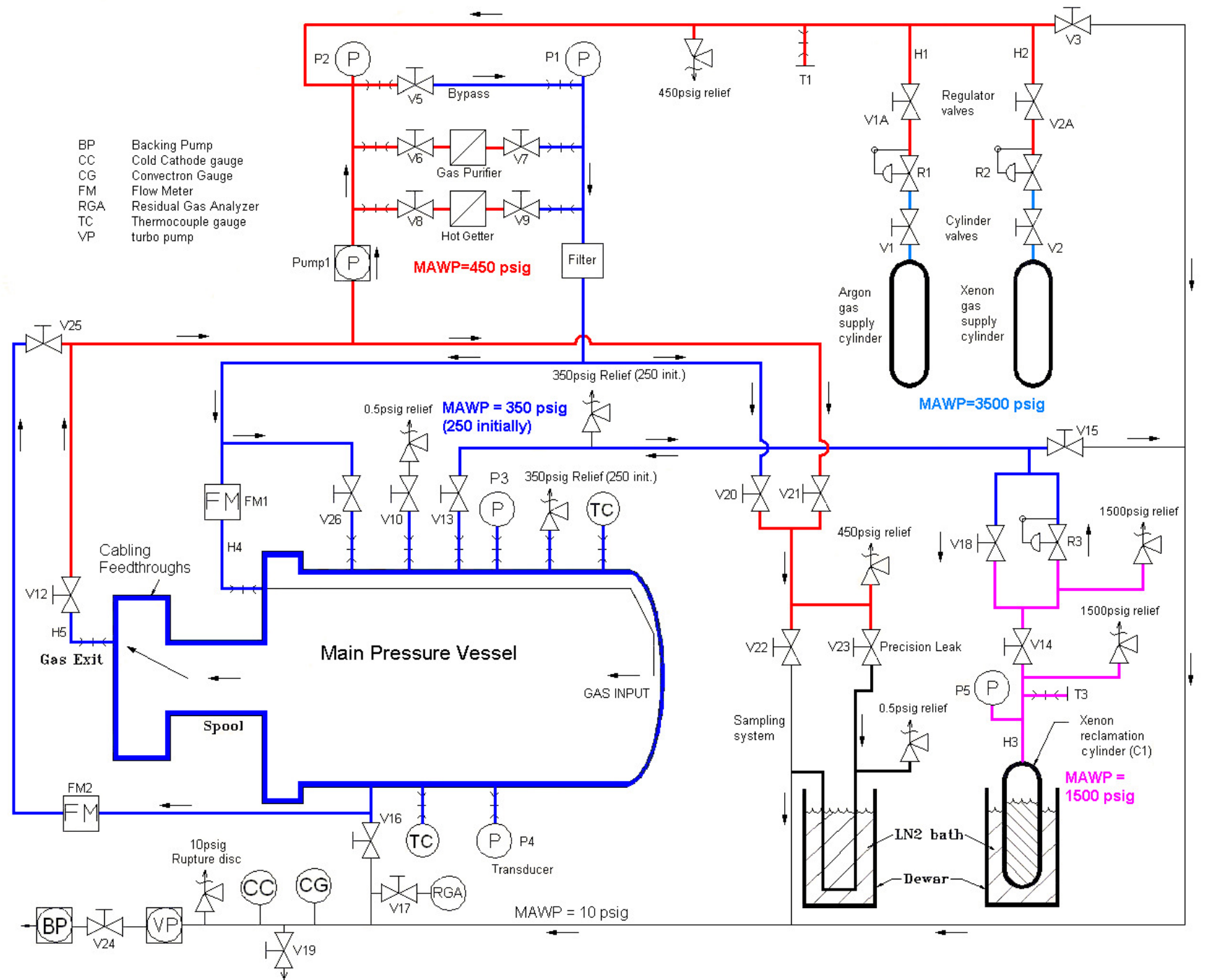}
  \caption{\label{gas_system_fig} \textbf{Gas system schematic (simplified):} The bottom left of the diagram shows the main 
pressure vessel with its connections to the vacuum, fill, recirculation/purification, sampling and reclamation components.}
\end{figure}

In Figure \ref{gas_system_fig} the complete gas system is shown. Besides the recirculation pump and the heated getter
the system includes a vacuum system with a roughing pump (make TriboVacDry), a turbomolecular pump 
(make Leybold-Heareus) and Pirani and ion vacuum gauges, a reclamation cylinder where xenon is stored
after it is cryogenically removed from the main pressure vessel, an argon purge system, a gas sampling system with a
precision leak valve and a residual gas analyzer (SRS model RGA100) and a room temperature secondary getter 
(SAES model MicroTorr MCP190).

A set of pressure relief valves (with different settings for the various parts of the pressure rated system) and a burst disk in the vacuum 
system protect the equipment and personnel from overpressure hazards.

The typical gas cycle during normal operation of the TPC consists of a vacuum step to  ~10$^{-5}$ torr, 
followed by an argon purge and recirculation at 1 atm, a second vacuum step to remove the argon and then
the xenon fill from the reclamation cylinder.  After the fill, the recirculation and purification is started, monitored 
with gas flow meters (models Sierra Smart-Trak and Omega FM1700) and controlled with a variac that powers the
recirculation pump to set the recirculation flow. 

\subsection{Pressure Vessel and Feedthrough Ports}

The main pressure vessel is an 8.7 liter stainless steel cylindrical shell of 20 cm diameter and 33.5 cm length with an 
elliptical head on one end and a custom gasketed Conflat flange (main flange) on the other. Half inch VCR ports on the side of the vessel are used 
for gas fill, recirculation flow, pressure and temperature gauges and pressure relief valves. A 5.9 cm diameter (ID) port is used for vacuum 
pumping. On the main flange there are small ports with commercial high voltage feedthroughs (rated to 20 kV and 17 atm) with additional custom 
PTFE sleeves on the inside to increase the breakdown voltage when operating with high pressure xenon. A larger central port is used to connect to 
an octagon-shaped vessel through a long tube with an internal source reentrance tube with a 2 mm thick endcap. Signal and high-voltage coaxial 
cables from the (in-vessel) PMT array pass through the axial extension tube and connect with 32-pin feedthroughs on the side ports of the octagonal 
vessel. 
     
\subsection{TPC}
\begin{figure}[!htb]
\centering
  \includegraphics[scale=0.15, trim = 50mm 0mm 0mm 0mm, clip]{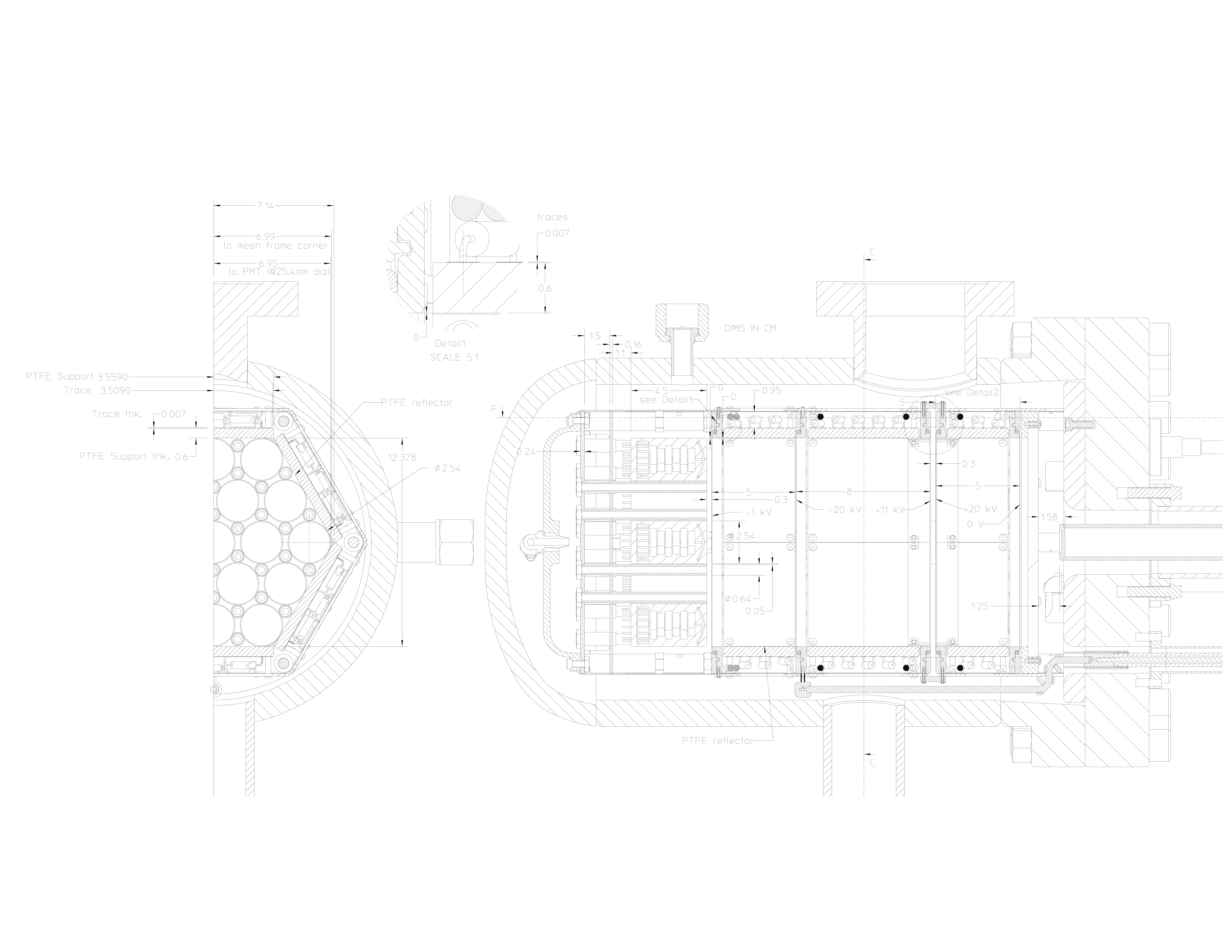}
  \caption{\label{tpc_schematic_fig} \textbf{Cutaway schematic of the TPC.}}
\end{figure}

 The field configuration in the TPC is established by five stainless steel meshes with 88\% open area at a $z$ position of 
0.5 cm (cathode buffer or PMT mesh), 5.5 cm (cathode or drift start mesh), 13.5 cm (field transition or EL-start mesh), 14.0 cm (anode or EL-end mesh) and 19.0 cm (anode buffer or ground mesh) from the PMT windows. Electroluminescence occurs between 13.5 and 14.0 cm. The meshes are supported
and kept tense by stainless steel frames made out of two parts and tensioning screws on the perimeter.   
The TPC side walls, made out of 18 individual rectangular assemblies 7.1 cm wide (and 5 and 8 cm length) connecting adjacent 
meshes (except around the 0.5 cm EL gap), serve the dual purpose of light cage and field cage.
Each side wall assembly is made of a 0.6 cm thick PTFE panel and a ceramic support panel. The FTFE panels are bare on the 
side facing the active volume and have copper stripes  parallel to the mesh planes every 0.6 cm on the other side. The bare PTFE 
serves as reflector for the VUV light (a 40-50\% Lambertian reflectivity was measured in \cite{Silva:2010}) to increase 
the light collection efficiency. Adjacent copper stripes are linked with 100 M$\Omega$ resistors to grade the potential and 
produce a uniform electric field. The ceramic support panels are connected, mechanically and electrically, to the outer 
perimeter of the mesh support frames and to the first and last copper stripes on their corresponding PTFE panel. High voltage 
connections to establish the TPC fields (HHV) are made directly to the mesh frames.          

Six short PEEK rods going through holes on the mesh frames' vertices secure the anode and anode buffer meshes to the main 
vessel flange. Three PEEK c-clamps (on alternate sides of the hexagon) with grooves to prevent HV surface breakdown attach, 
in turn, the anode mesh to the field transition mesh maintaining a gas gap of 0.5 cm. Finally, six PEEK rods support the field transition, cathode 
and cathode buffer meshes as well as the 19-PMT array. Mechanical tolerances obtained are better than 1 mm throughout the 
TPC geometry with better tolerances on the electroluminescent gap between 13.5 and 14.0 cm.    

In the initial implementation of the NEXT-DBDM TPC longer PEEK rods supported the entire assembly going through holes in the anode and field transition mesh 
frames with PEEK spacers to maintain the EL gap. High voltage breakdown in the form of sparks across the gap during HV conditioning and at random 
times thereafter produced conductive paths on the insulator surface, requiring time consuming repair. The c-clamp solution described above made the 
TPC completely resilient to the unavoidable occasional sparking.

Clean gas from the purification system flows through an internal tube to an enclosed volume behind the PMTs and reaches the active volume through 
small dedicated PEEK tubes between the PMTs to exit the TPC through a port on the octagonal vessel at the end of the extension tube.

A wide range of HHV voltages were used to investigate the TPC performance dependence on drift electric field and on the {\it E/P} in the EL region: 
cathode voltages from -4 to -13 kV, field transition voltages from -1 to -10 kV and anode voltages from 1 to 13 kV.

\subsection{PMTs, Readout and Data Processing}

The PMT array is composed of nineteen 2.54 cm diameter Hamamatsu 7378A PMTs.  These PMTs, with fused silica windows, have a 
quantum efficiency of $\sim$15\% for 172 nm xenon light. The PMTs were individually pressure tested to 19 atm (absolute) and 
no mechanical or performance failures were observed. The HV, typically about -1 kV, is individually set to get a $\sim10^6$ 
gain. The bases for the PMT array are implemented in a single hexagonal PCB board with surface mount components and with pin 
sockets that provide both mechanical support for the PMTs as well as the necessary electrical connections. The base has a total resistance close to
1 M$\Omega$, thus power dissipation is about 1 Watt per PMT. 1 $\mu$F capacitors connected to the last dynode stages 
keep the gain constant during long EL light pulses. 200 pF capacitors connected from the PMT anodes to ground stretch the pulse waveform such that all photoelectrons are properly sampled in the 100 MHz digitizers.      

PMT anode signals travel through $\sim$1 m long PTFE coaxial cables to the 32-pin feedthoughs and then to 8-channel 
Phillips-777 amplifiers set to a gain of 40. The amplified signals are then stretched and attenuated in a passive RCR 
circuit to reduce high frequency noise and to match the input range of the SIS3302 16-bit digitizers that sample the 
individual PMT signals at 100 mega-samples per second. PMT waveform data, typically 16,384 samples or 163.84 $\mu$s, are 
readout through an SIS3150 USB to VME interface to a Linux server for processing, analysis and storage. The overall system 
noise is such that individual photoelectrons can be detected and the instantaneous (10 ns) dynamic range per PMT is 1-to-200 
photoelectrons.  Custom-written DAQ software is used to control the data acquisition. 

The trigger is designed to identify S2 EL pulses which have much larger area than the S1 pulses. The signal from a single PMT is
integrated by a 12$\mu$s peaking-time shaping amplifier and then sent to a discriminator. This scheme effectively provides an 
S2 energy threshold. The trigger is used as a stop signal (with a stop delay) for the digitizers such that 
there are, typically, 80 $\mu$s of waveform data preceding the first S2 pulse and 80 $\mu$s after. Since the maximum drift 
time is about 80 $\mu$s, this permits the search for the S1 pulse in the offline analysis while ensuring that all the S2 
light from one event is measured.      

After a block of 512 events is collected with about 100 MBytes in raw DAQ data format, an automated process unpacks the data, 
applies calibration constants to the individual PMTs and executes the analysis code based on ROOT \cite{ROOT} and 
FMWK \cite{FMWK} that finds S1 and S2 pulses and computes energies, times and positions and outputs the results in 1 
MBytes ROOT data trees.

\subsection{Controls and Monitoring}

All system controls  (except the single PMT HV power supply), such as HHV voltage settings and current limits and recirculation flow,  are done 
manually.  The three HHV voltages and currents, the PMT HV power supply voltage, the pressure inside the TPC vessel, the temperature at two points 
inside the TPC and the room temperature are automatically and continuously recorded for monitoring and to aid in offline data analysis. 
The recirculation flow is recorded manually.

\section{Data Analysis}

\begin{figure}[!htb]
  \centering
  \includegraphics[scale=0.7]{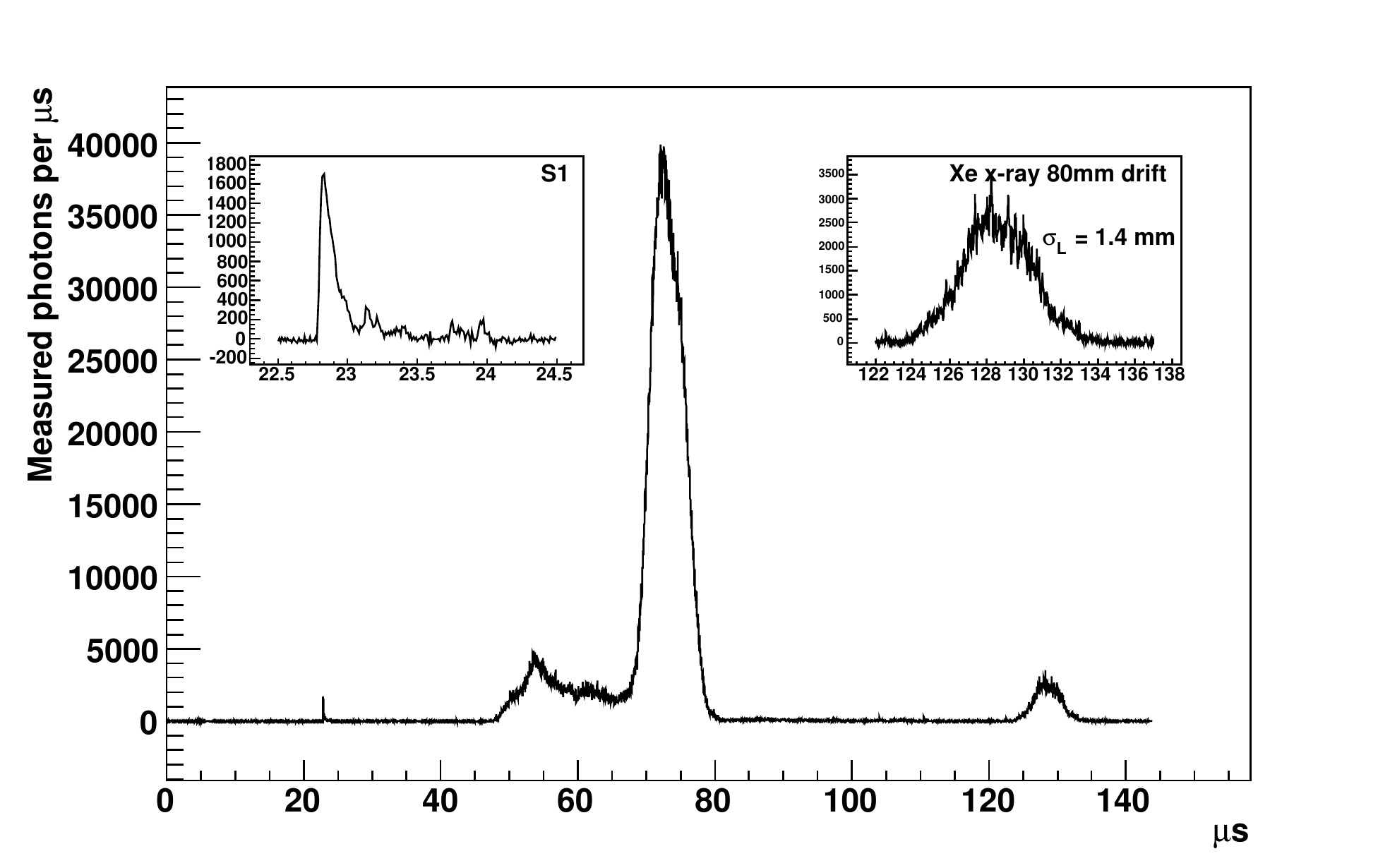}
  \caption{\label{waveform_long_drift} \textbf{Typical full energy 662 keV gamma ray event waveform:} the sum of the previously calibrated waveforms of the 19 
  photomultipliers is shown. The S1 pulse (shown in detail in the left inset) due to xenon scintillation is short and with O(200) 
  measured photons. Two S2 pulses caused by electroluminescence of xenon from ionization electrons follow. The first with $\sim$270,000 measured 
  photons is likely due to a 630 keV electron from the photoelectric interaction of the 662 keV gamma ray; the pulse structure reflects the 
  ionization density of the track along its $\sim$2.5 cm long projection on the drift (z) direction. The second (shown in detail in the right inset) 
  with $\sim$12,000 measured photons is likely due to the interaction, a few cm away, of a 30 keV xenon X-ray following the photoelectic process; 
  since the 30 keV energy deposition is nearly point-like the pulse shape is gaussian with a $\sigma_L$ of 1.4 mm set by the longitudinal diffusion 
  of the electrons during the $\sim$8 cm drift. This event is from a data run taken at 10 atm with a 0.16 kV/cm drift electric field and an {\it E/P} of 
  2.1 kV/(cm atm) in the EL region. }
\end{figure}

\begin{figure}[!htb]
  \centering
  \includegraphics[scale=0.6]{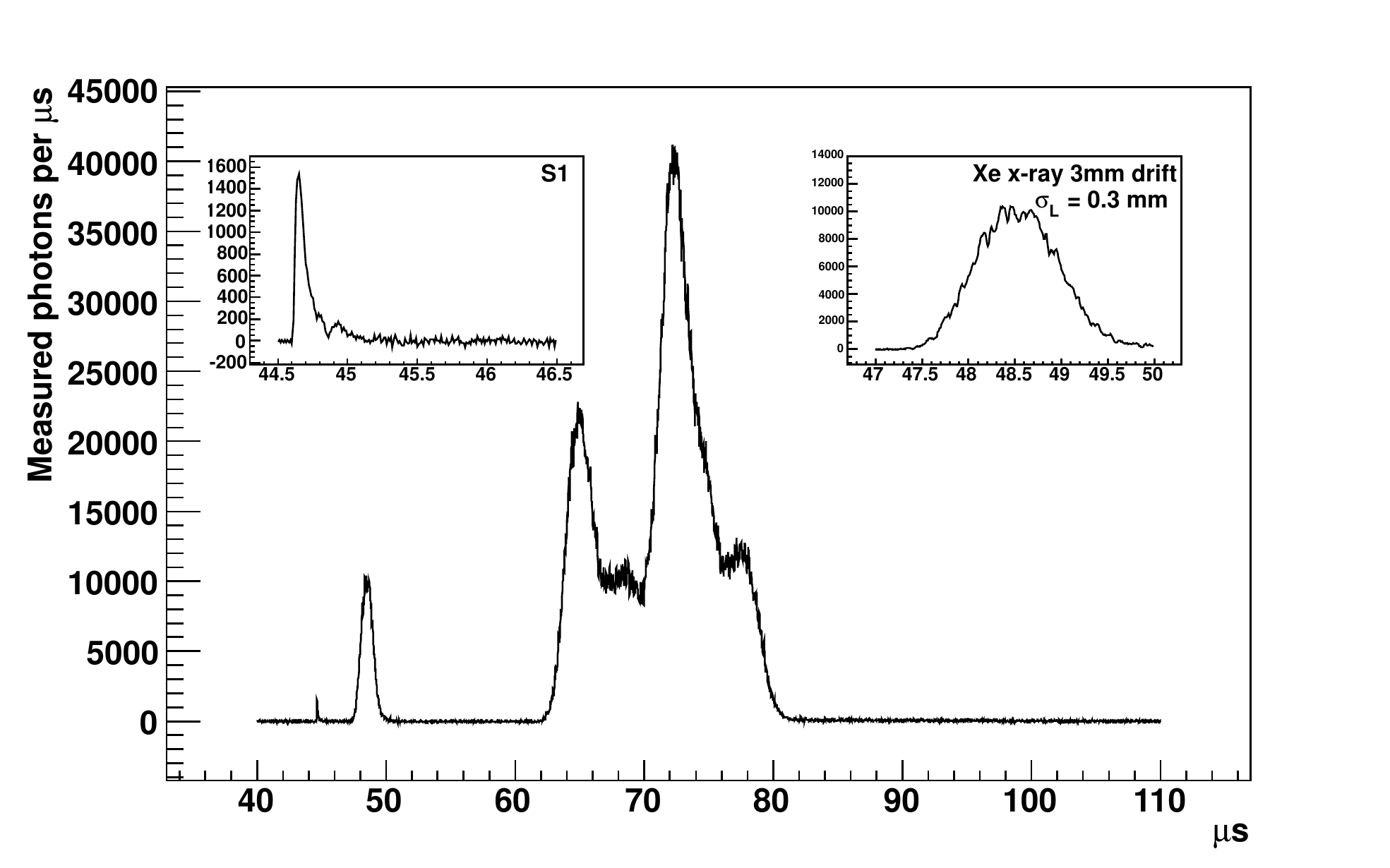}
  \caption{\label{waveform_short_drift} \textbf{A second typical full energy event:} In this event from the same data run as in Fig. \ref{waveform_long_drift} the xenon X-ray interacted close to the EL region. The ionization electrons from it drifted for just 3 mm and underwent a correspondingly small longitudinal diffusion $\sigma_L$ of 0.3 mm. }
\end{figure}

In the first step of the data analysis the individual PMT waveforms are changed to photoelectron units using calibration constants 
from dedicated low occupancy single photoelectron runs with short LED light pulses. A sum waveform is then computed from a sample-by-sample 
addition of the 19 PMTs' waveforms and the baseline of the sum waveform is obtained. A search for S1 and S2 pulses that cross a threshold follows. 
This threshold is determined from the value of the baseline noise. Pulses are defined as S1 candidates if they are less than 500 ns wide and have a maximum
rise time (defined as the time from the start of the pulse to the location of the peak amplitude) of 100 ns. All other 
pulses are considered S2 candidates. Individual pulses are integrated and the largest S1 candidate is assumed to be the event's start time 
and the others discarded. All S2 pulses that follow the chosen S1 are considered part of the event while the ones preceding it are discarded. 
An event is considered valid if it has an S1 pulse with at least one associated S2 pulse. Figures \ref{waveform_long_drift} and 
\ref{waveform_short_drift} show typical valid events from 662 keV $^{137}$Cs gamma ray interactions. 
    
During the automatic analysis of the data, the electron attachment lifetime of the gas is not yet known. In order to enable the offline correction of these charge losses the ten first moments of the S2 charge distribution  $\int q(t) t^n dt$ with n=0,1...,9 are calculated where $t$ is the time interval since the S1 start time pulse. 

An average $x-y$ position for the event is calculated from an S2-charge weighted average of the PMT positions. The event is also time split into equal-charge slices. For each slice an average $x-y$ position is computed from the PMT light distribution. This provides $x-y-z$ positions for well separated energy depositions such as from Compton scatters or from X-rays following photoelectric interactions.

\section{Results}

\subsection{PMT Performance}

As shown in formulas \ref{eq:resolution} and \ref{eq:G} the energy resolution achievable in an electroluminescent TPC depends on the precision with which individual photons can be measured. System noise and the variance from the avalanche multiplication in the PMT (mostly from the first dynode stages) contribute to the spread of the charge measured for photoelectrons. We found, however, that the charge fluctuations due to PMT after-pulsing are the dominant factor in the photoelectron charge resolution. After-pulsing caused by the occasional ionization of residual gas molecules in the PMT vacuum volume can produce delayed pulses with large charges (10-20 photoelectron charges are not uncommon). Dedicated LED data runs with light pulses less than 100 nsec long (first afterpulse peaks from hydrogen and helium ions appear at 200 and 300 nsec respectively for these small PMTs) were thus taken to assess the charge-variance for each PMT.  
The typical value was found to be $\sigma(Q)/Q=1.2$ thus $\sigma_\mathrm{pd}^2$=1.44  with small PMT-to-PMT variations.

\subsection{Position Measurement}
\begin{figure}[!htb]
  \centering
  \includegraphics[scale=0.4]{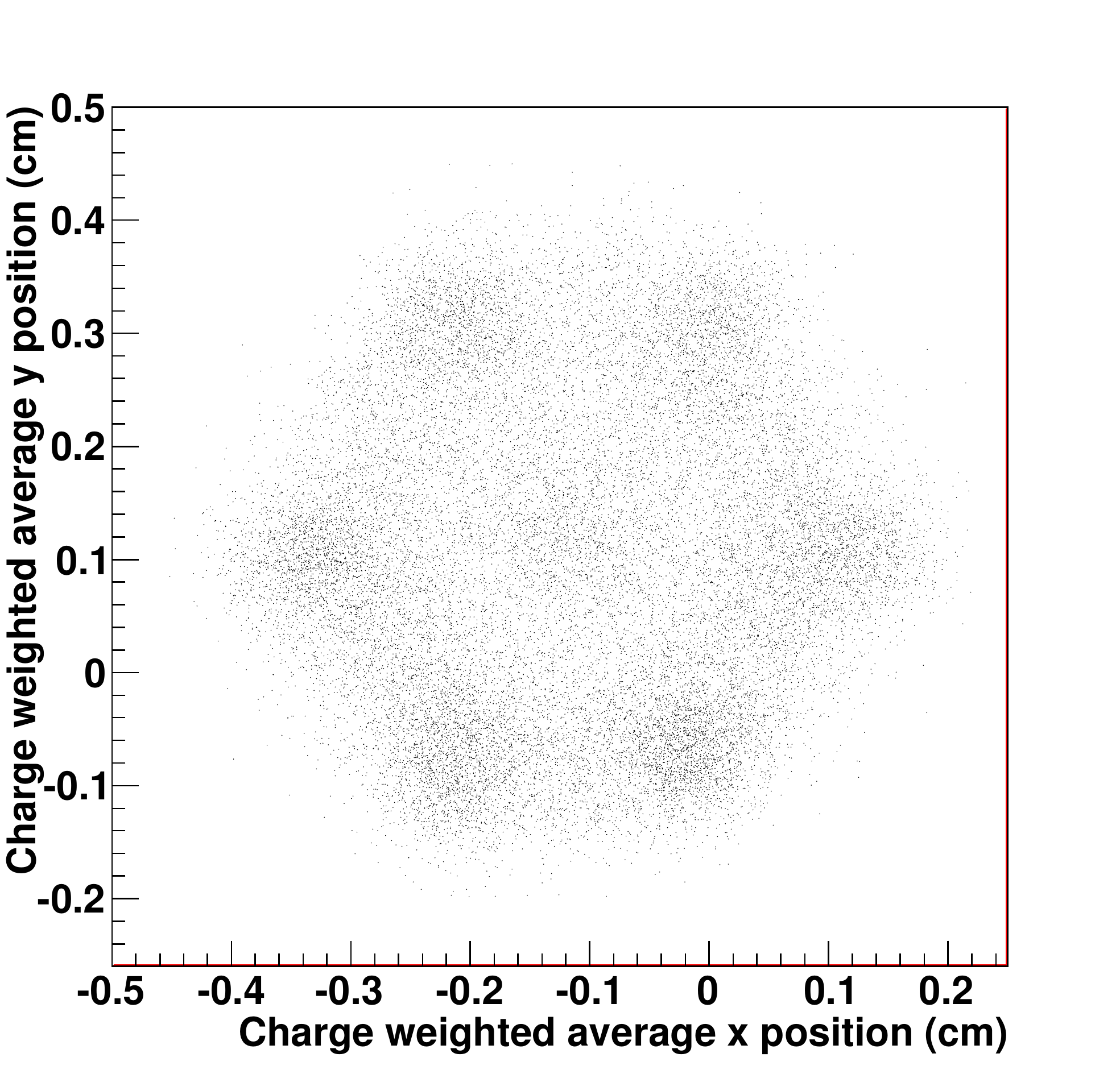}
  \caption{\label{position_calibration} \textbf{Position reconstruction:} The charge-weighted average of the 19 PMT positions is used for $x-y$ reconstruction. Events with energy depositions between 50 and 100 keV were selected; at these energies ionization tracks extend for just a few mm and produce enough light to reconstruct with sufficient position resolution. The edges of the hexagonal area correspond to the TPC cross section. Due to the spatial uniformity of the PMT plane illumination a scaling of factor of $\sim$30 (not applied here) is needed to obtain true $x-y$ positions from these charge-weighted averages. These data were taken with the $^{137}$Cs source, at 10 atm and with {\it E/P} of 2 kV/(cm atm) in the EL region.}
\end{figure}

The NEXT-DBDM TPC configuration with the PMT array 13.5 cm  from the EL region does not permit detailed track reconstruction in the $x-y$ plane. Still, the position reconstruction achievable allows the fiducialization of pulses to select events/pulses within regions of the TPC with uniform light collection efficiencies. The position reconstruction for isolated 50-100 keV energy depositions shown in Fig. \ref{position_calibration} displays the hexagonal boundary of the TPC. A scaling factor of $\sim$30 and an $x-y$ offset are needed to convert charge-weighted average $x$ and $y$ positions to true TPC $(x,y)$ coordinates. The non-uniformity observed in Fig. \ref{position_calibration} is due to the inaccuracies of the simple charge-weighting algorithm. A detailed Monte Carlo simulation of the experimental setup, including the EL light propagation and the optical properties of the detector components shows the same non-uniformities, a nearly linear correspondence between true and reconstructed positions and the same $\sim$30 scaling factor between them. The $x-y$ offset, on the other hand, is unique to the TPC data and is due to small PMT-to-PMT differences in quantum efficiency and gain (not fully removed by the dedicated calibration).  

\begin{figure}[!htb]
  \centering
  \includegraphics[scale=0.6]{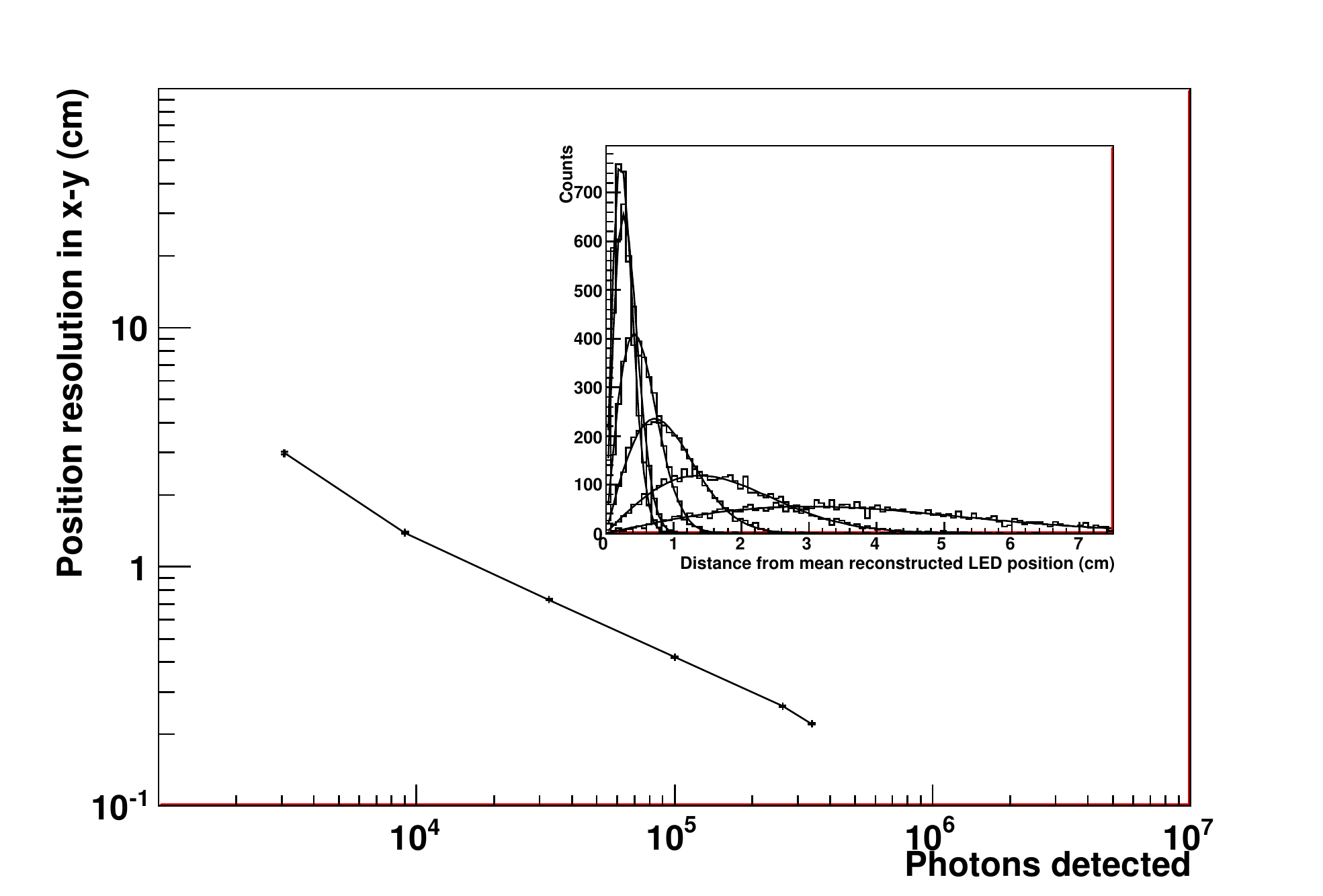}
  \caption{\label{led_position_res} \textbf{Position resolution for LED light pulses:} The position resolution in the $x-y$ plane as a function of the number of detected photons is shown. The resolution is derived from data from light pulses from a 378 nm LED located behind the PTFE backplane, which diffuses its light, and illuminates the PMT plane from a distance of about 19 cm. The inset shows the radial distributions (and fits) from which the individual resolution values were obtained.}
\end{figure}

Several data runs were taken with LED light pulses of various intensities to determine the position resolution for point depositions of charge as a function of the amount of light produced and detected. Figure \ref{led_position_res} shows the obtained position resolutions which approximately follow the expected $1/\sqrt{N}$ dependency on the photon statistics.  Point-like depositions of charge, such as those from xenon X-rays, can thus be reconstructed in $x-y$ with better than 1 cm resolution if at least 10,000 EL photons are detected from them.  

\subsection{Energy Measurement}

\begin{figure}[!htb]
  \centering
  \includegraphics[scale=0.7]{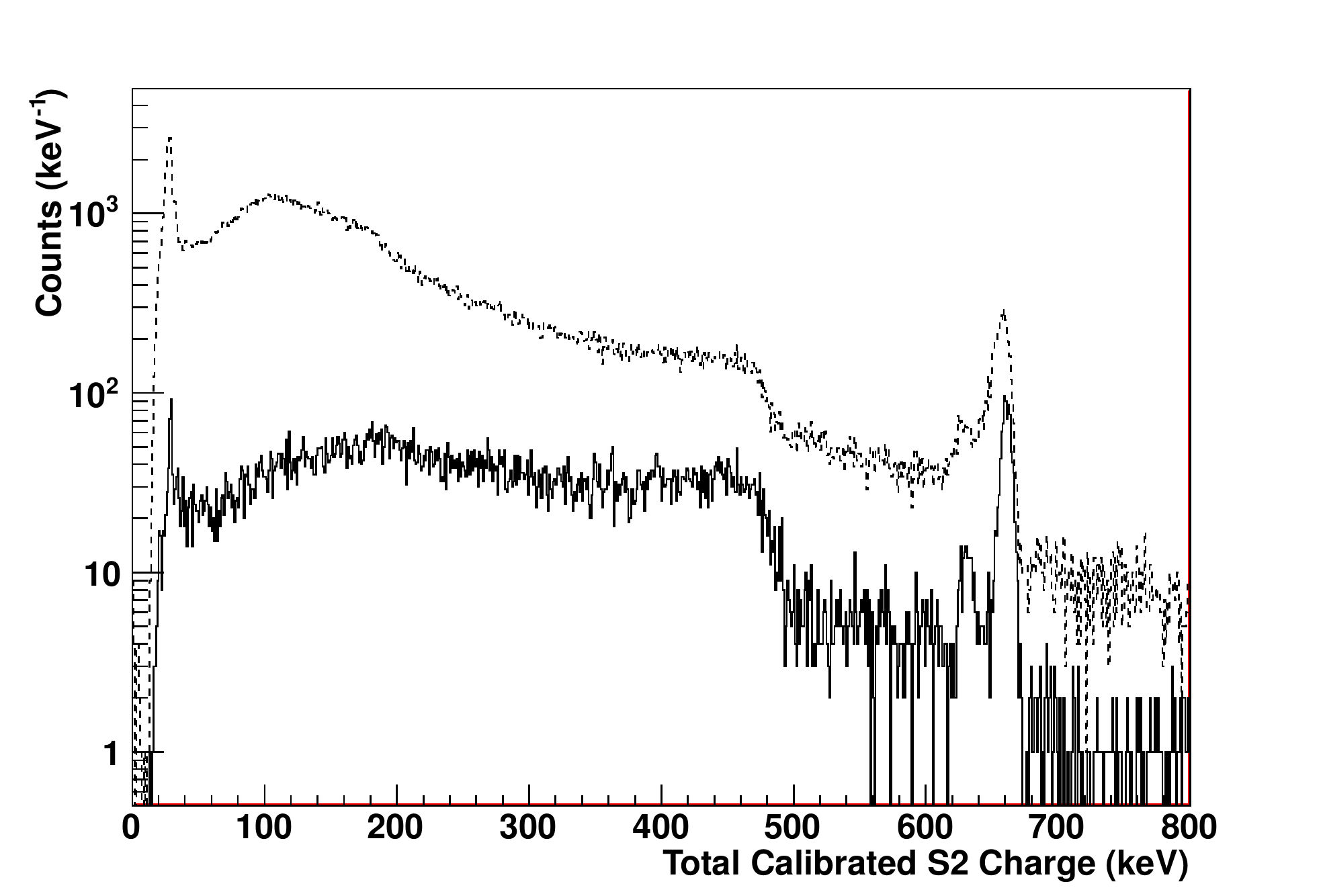}
  \caption{\label{full_spectrum} \textbf{Energy spectrum for $^{137}$Cs 662 keV gamma rays:} The dashed line shows the calibrated spectrum 
  (using the full energy 662 keV peak alone) from a data run taken at 10 atm with a 0.3 kV/cm drift electric field and an {\it E/P} of 2 kV/(cm atm) 
  in the EL region. The 1 mCi source is strongly collimated. Gamma rays enter the TPC along the drift axis (z). Below the full energy peak 
  at 662 keV, the smaller X-ray escape peak at $\sim$630 keV, the Compton edge at 477 keV and the xenon X-ray peak at $\sim$30 keV can be seen.  
  The solid line is the same spectrum with a requirement that the reconstructed average position of an event be   less than $\sim$1.2 cm from the 
  TPC axis. The small feature near 184 keV is due to Compton backscatters. The no-source background spectrum (not shown) has a broad peak near 100 
  keV.}
\end{figure}

The energy measurement is derived from the summed waveform (in photons detected) of all S2 pulses following the event S1 pulse. Figure 
\ref{full_spectrum} shows a calibrated spectrum obtained from the  $^{137}$Cs source collimated with the gamma rays entering the chamber along its 
central axis. The full energy peak events in this data run have $\sim$270,000 detected photons which, for $W_\mathrm{i}$=24.8 eV, corresponds to about 10.1 
photons detected per ionization (secondary) electron. Using the nominal EL gain $\eta$=846 from Eq. \ref{eq:elgain} the total number of photons 
produced is $846\cdot662,000/24.8\simeq$ 22.6 million. From this, an S2-photon detection efficiency of 1.2\% is deduced.

While the spectrum was calibrated only using the 662 keV line, the xenon X-rays peak appears at the correct energy (two strongest X-ray lines are at 29.782 keV and 29.461 keV \cite{LBLTOI}) confirming the expected linearity of the energy measurement in the interval of interest here. 
 
\begin{figure}[!htb]
  \centering
  \includegraphics[scale=0.6]{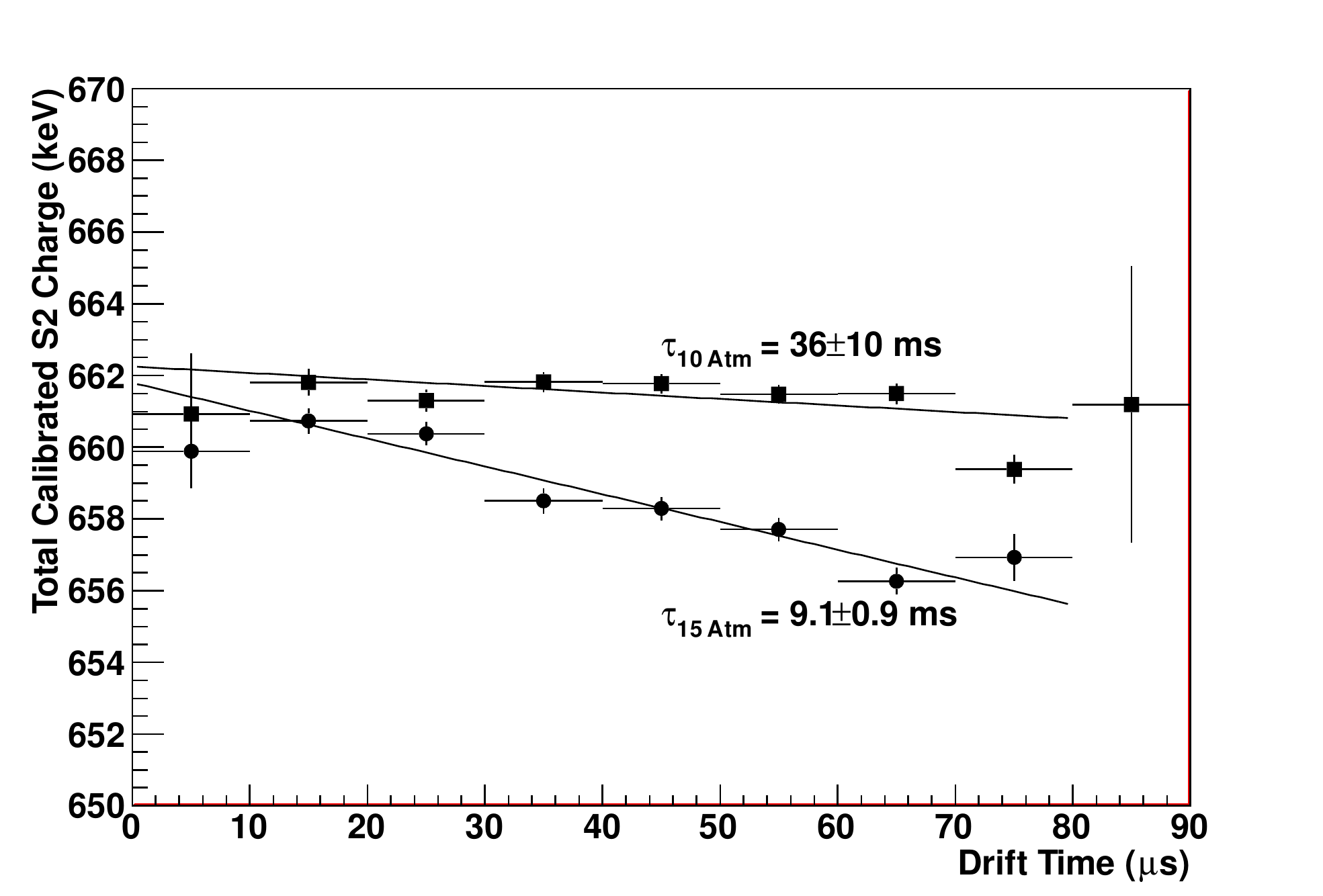}
  \caption{\label{attachment} \textbf{Electron attachment losses:} The average calibrated S2 charge versus the charge-weighted drift time is shown 
  for full energy events. Attachment lifetimes are obtained from the exponential fits. The 10 atm data was taken with a 0.19 kV/cm field 
  in the drift region and the 15 atm data with 0.59 kV/cm. }
\end{figure}

Attachment losses of drifting secondary electrons can be assessed from the full energy peak position as a function of the drift time of the events. 
Figure \ref{attachment} shows typical data for 10 and 15 atm with excellent electron lifetimes of 36 and 9 milliseconds, respectively. Less than 
1\% of the electrons are lost to attachment for the maximum 8 cm drift length. Assuming that the main source of electron attachment is due to 
residual O$_2$, the measured electron lifetime at 10 atm corresponds to a residual oxygen component in the 2-4 part per billion range \cite{Magboltz}. The lower 
lifetime at 15 atm is due to the quadratic pressure dependence of the 3-body attachment process rate.
 
\begin{figure}[!htb]
  \centering
  \includegraphics[scale=0.6]{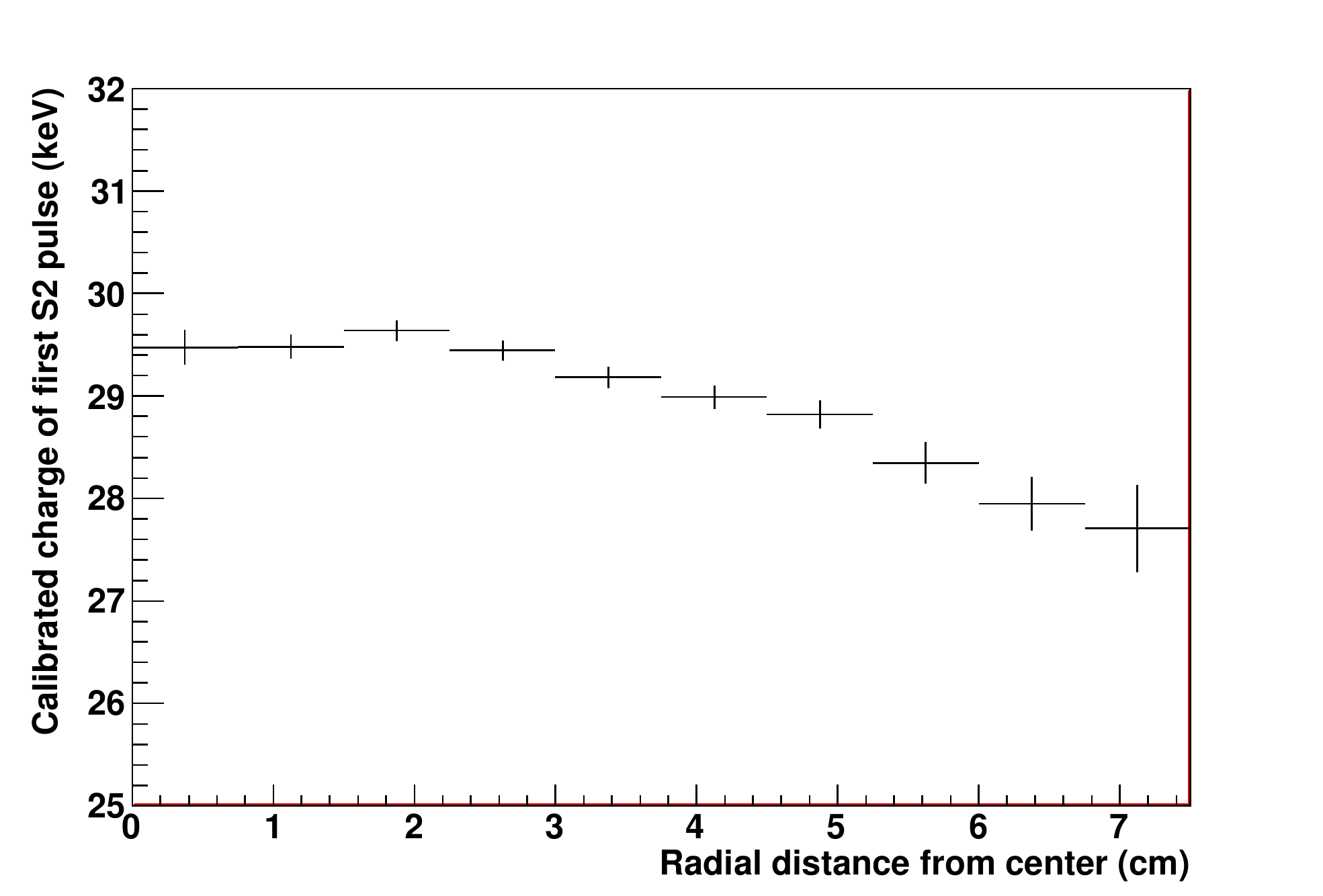}
  \caption{\label{xray_radial_dependence} \textbf{Radial dependence of energy response:} Xenon X-ray energy depositions were selected from 
  full-energy 662 keV events for which the first S2 pulse has energy in the 20-40 keV range. The reduced response at larger radial positions 
  (also observed in a detailed Monte Carlo simulation of the TPC) is due to geometric and optical effects of the EL light collection efficiency. 
  These data were taken at 10 atm with a 1.0 kV/cm field in the drift region and 2.68 kV/(cm atm) in the EL region. }
\end{figure}

The S2 photon collection efficiency is expected to vary somewhat as a function of the $x-y$ position because of solid angle and chamber optics 
(wall reflectivities, etc). A sample of xenon X-ray energy depositions was used to measure the change in photon detection efficiency as a function 
of the radial position. Figure \ref{xray_radial_dependence} shows this dependence derived from a run in which individual X-rays were reconstructed 
with $\sim$0.8 cm resolution in $x-y$. For the purpose of this measurement X-ray depositions can be considered point-like because 30 keV electron 
tracks at these pressures extend typically for less than half a millimeter \cite{Biagi:MIP}. A nearly flat response is seen up to 2 cm radius followed by a linear decline in response towards the outer edge of the TPC where the collection is about 10\% lower than in the center. 

\begin{figure}[!htb]
  \centering
  \includegraphics[scale=0.6]{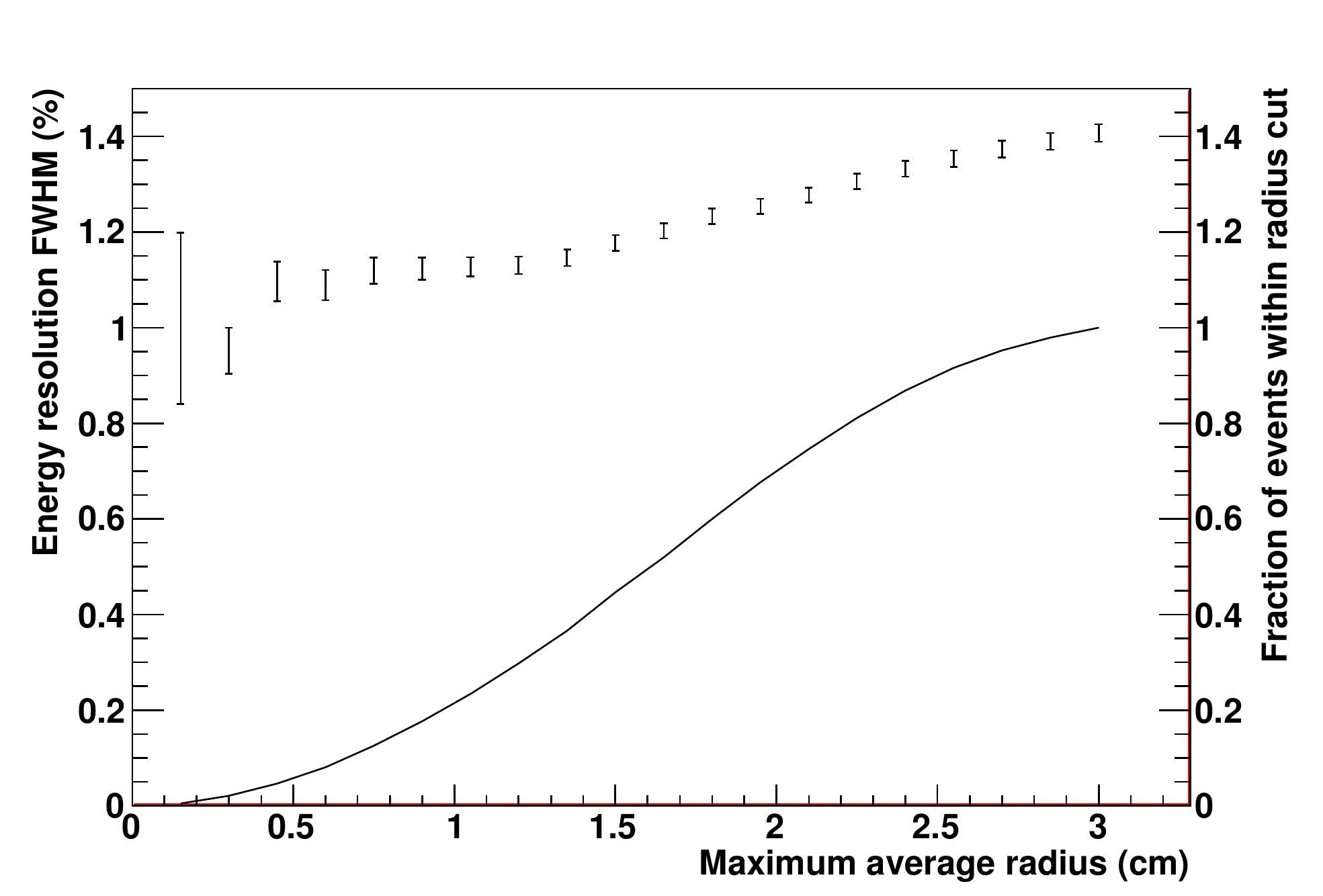}
  \caption{\label{cs137_resolution_rcut} \textbf{Energy resolution dependence on radial cut:} Full energy 662 keV events are chosen and, for each, 
  an average $x-y$ position is computed (and scaled to true TPC coordinates) using all of the S2 signal. The FWHM energy resolution (data points) is 
  obtained from a gaussian fit to the peak after placing a cut on the radial position (the abscissa). The radial cut efficiency (solid line and 
  right ordinate axis) is also shown. These data were taken at 10 atm with a 0.16 kV/cm field in the drift region and 2.1 kV/(cm atm) in the EL 
  region. }
\end{figure}

In the absence of detailed track imaging, the radial dependence of the detector response to point-like depositions affects the TPC energy resolution for extended tracks. Electron tracks from 662 keV gamma interactions at the pressures of interest here have 2-4 cm spans and nearly random shapes due to the large coulomb multiple scattering in xenon.  Figure \ref{cs137_resolution_rcut} shows the effect of a radial cut on the energy resolution of 662 keV depositions and its efficiency.  Since the collimated gamma rays enter the chamber along the main axis, events with a small average radial position have better energy resolution primarily because they span a smaller radial range.
    
\begin{figure}[!htb]
  \centering
  \includegraphics[scale=0.6]{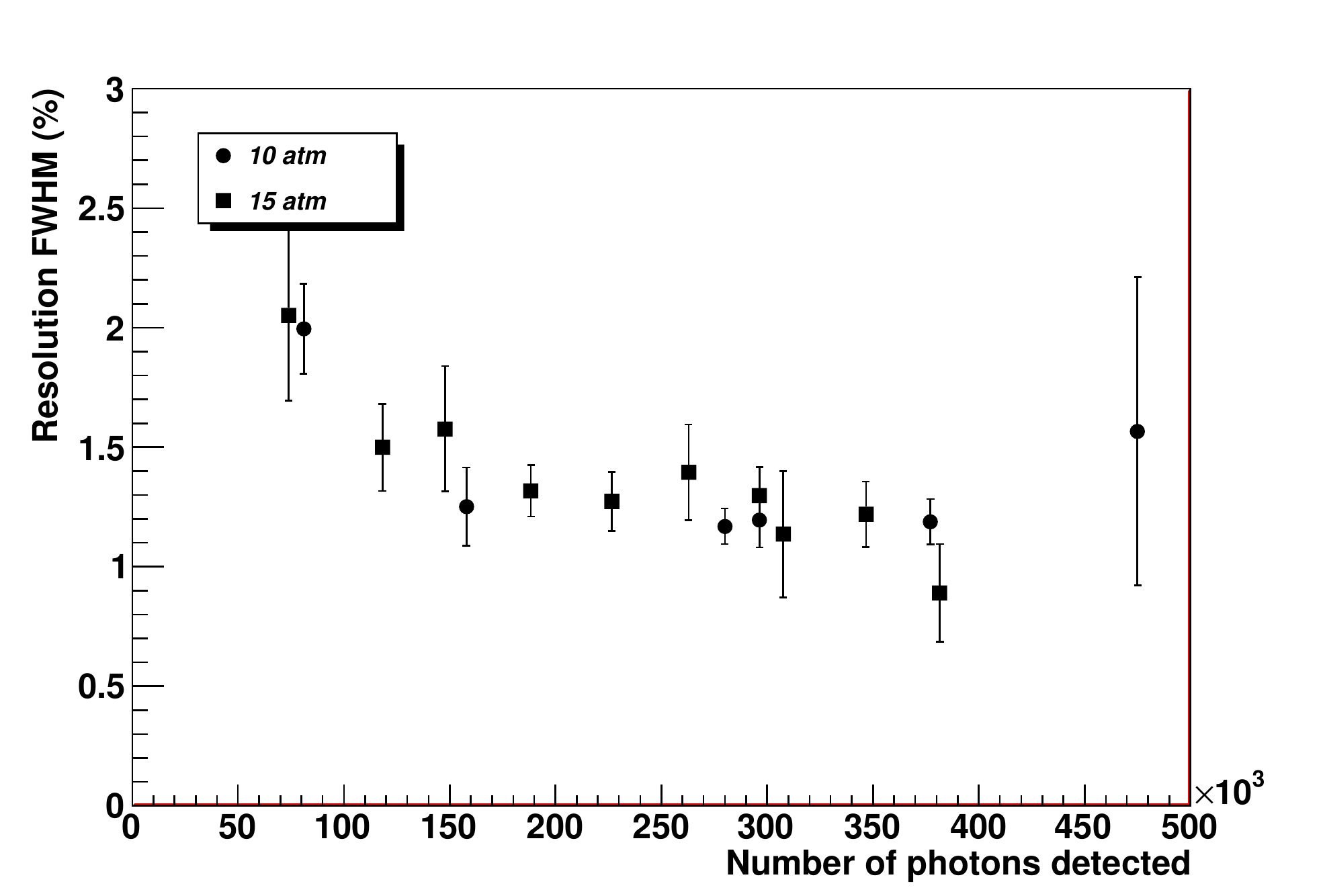}
  \caption{\label{el_dependence_resolution} \textbf{Energy resolution dependence on amount of light detected:} 662 keV full energy peak 
  resolutions are shown for data runs with different total S2 light yield (controlled primarily by the {\it E/P} in the EL region). Energy spectra 
  were obtained by selecting events reconstructed in the central 1 cm radius region and with an S1 signal with more than 100 measured photons 
  and not more than 600. A small electron drift-time-dependent attachment correction is applied. The expected $\sim$1/$\sqrt{N}$ improvement is 
  observed. These data were taken with drift fields of 0.5-0.6 kV/cm at 10 atm and 0.16-0.72 kV/cm at 15 atm. }
\end{figure}

Figure \ref{el_dependence_resolution} shows the improvement in the measured energy resolution for 662 keV depositions with increased 
EL light yield. As seen in Eq. \ref{eq:resolution} and \ref{eq:G} the G term dominates the resolution at low light yields ($n_\mathrm{pe}$ small) while 
the intrinsic Fano term is the asymptotic resolution in the large light yield limit.  

\begin{figure}[!htb]
  \centering
  \includegraphics[scale=0.6]{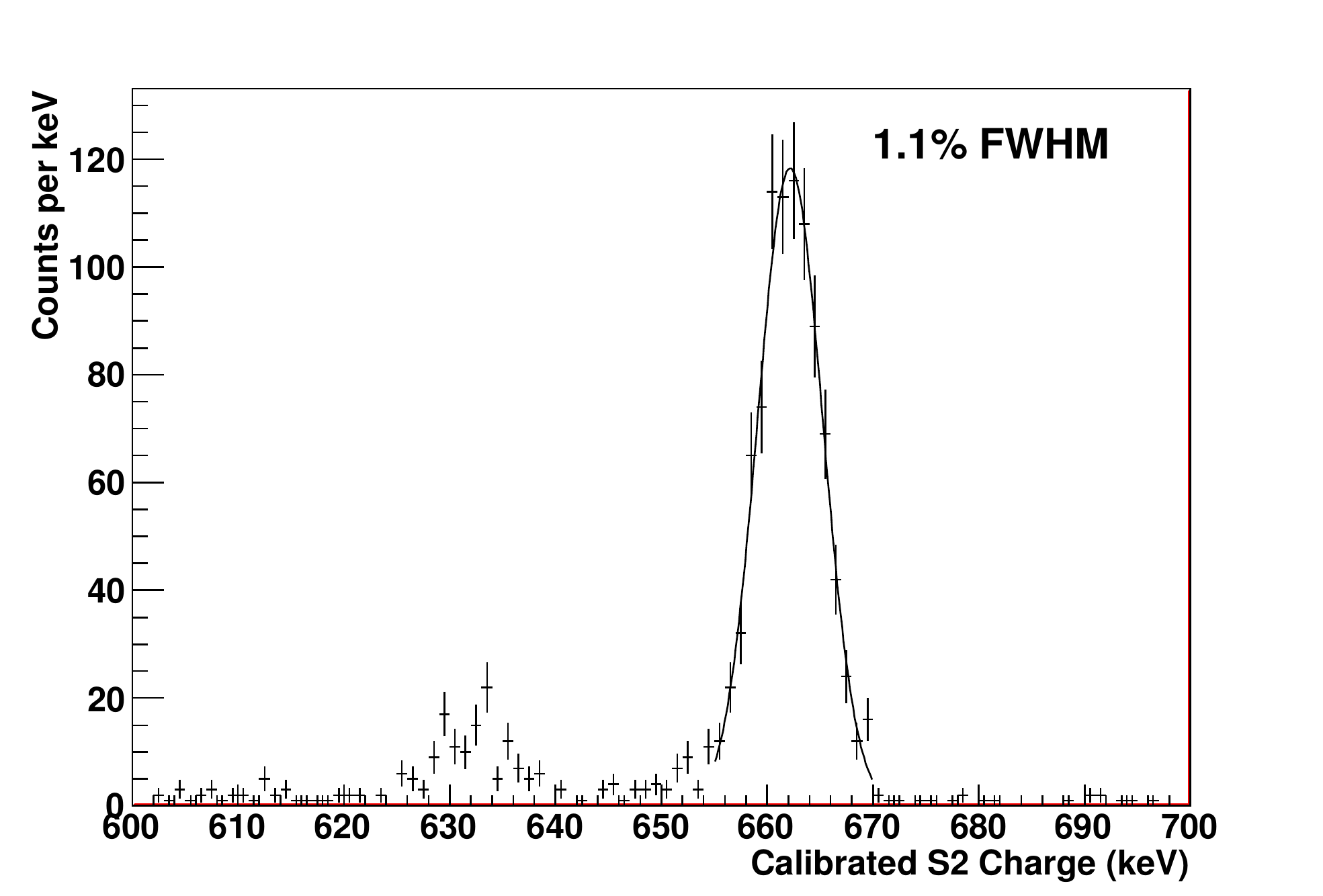}
  \caption{\label{cs137_resolution_10atm} \textbf{Energy resolution at 10 atm for 662 keV gamma rays:}  These data were taken at 10.1 atm with a 
  0.16 kV/cm field in the drift region and 2.08 kV/(cm atm) in the EL region. If assumed to follow a 1/$\sqrt{N}$ dependence this resolution 
  extrapolates to 0.57\% at $Q_{\beta\beta}$=2.459 MeV.}
\end{figure}

\begin{figure}[!htb]
  \centering
  \includegraphics[scale=0.6]{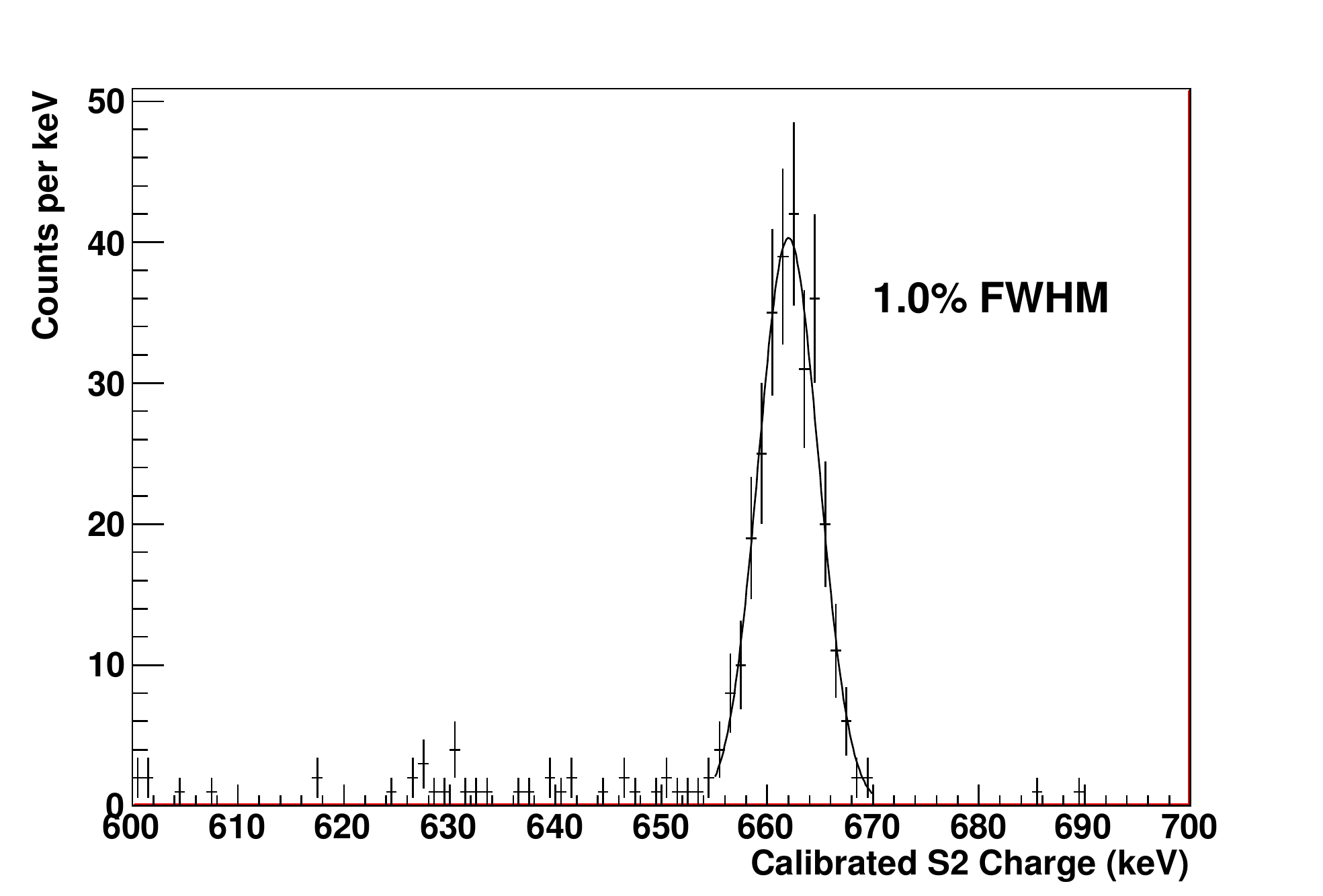}
  \caption{\label{cs137_resolution_15atm} \textbf{Energy resolution at 15 atm for 662 keV gamma rays:} A 1.0\% FWHM energy resolution was obtained 
  for events reconstructed in the central 0.75 cm radius region. The attachment losses correction with $\tau$ = 9.0 ms was applied. A PMT with a 
  clear time varying response was removed from the measurement.  These data were taken at 15.1 atm with a 0.59 kV/cm field in the drift region and 
  1.87 kV/(cm atm) in the EL region. }
\end{figure}

In Fig. \ref{cs137_resolution_10atm} the energy spectrum in the 662 keV full energy region obtained at 10 atm is shown. A 1.1\% FWHM energy 
resolution was obtained for events reconstructed in the central 0.6 cm radius region. A small drift-time dependent correction for attachment 
losses with $\tau$ = 13.9 ms was applied. The xenon X-ray escape peak is clearly visible $\sim$30 keV below the main peak. For the spectrum 
taken at 15 atm, shown in Fig. \ref{cs137_resolution_15atm}, a 1\% FWHM resolution was obtained. At this higher pressure the xenon 
X-ray is less likely to escape from the active volume and the escape peak almost disappears. This resolution extrapolates to 0.52\% FWHM at 
$Q_{\beta\beta}$=2.459 MeV if the scaling follows a statistical 1/$\sqrt{E}$ dependence and no other systematic effect dominates. 

\begin{figure}[!htb]
  \centering
  \includegraphics[scale=0.6]{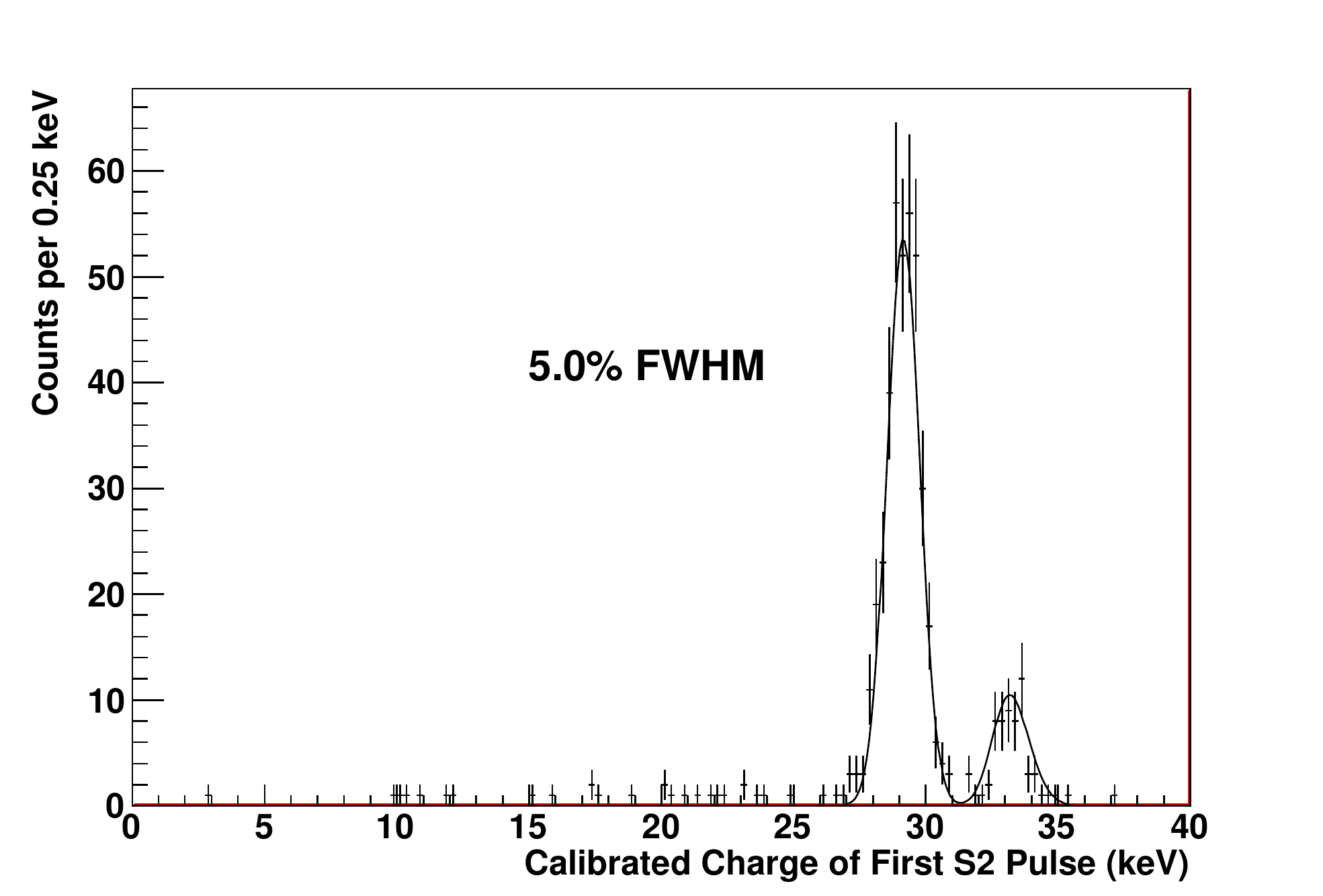}
  \caption{\label{xray_resolution} \textbf{Energy resolution at 10 atm for 30 keV xenon X-rays:} A 5\% FWHM energy resolution for 30 keV was 
  obtained These data were taken at 10.1 atm with a 1.03 kV/cm field in the drift region and 2.68 kV/(cm atm) in the EL region. }
\end{figure}

In order to study the EL TPC energy resolution at lower energies, full energy 662 keV events that had a well separated X-ray pulse reconstructed in 
the central 1.5 cm radius region were used.  The energy calibration was done on the 662 keV full energy peak and linearity with a zero intersect 
was assumed. Figure \ref{xray_resolution} shows the energy spectrum obtained at 10 atm with a 5\% FWHM resolution. The spectrum was fit to the sum 
of four gaussians with relative positions and intensities fixed to the strongest xenon X-ray lines \cite{LBLTOI}. The absolute position of the 
anchor peak and the peaks' widths (all assumed the same) were obtained from the fit. The anchor peak is at 29.1 keV, less than 2\% away from its 
nominal 29.6 keV value. 

\begin{figure}[!htb]
  \centering
  \includegraphics[scale=0.7]{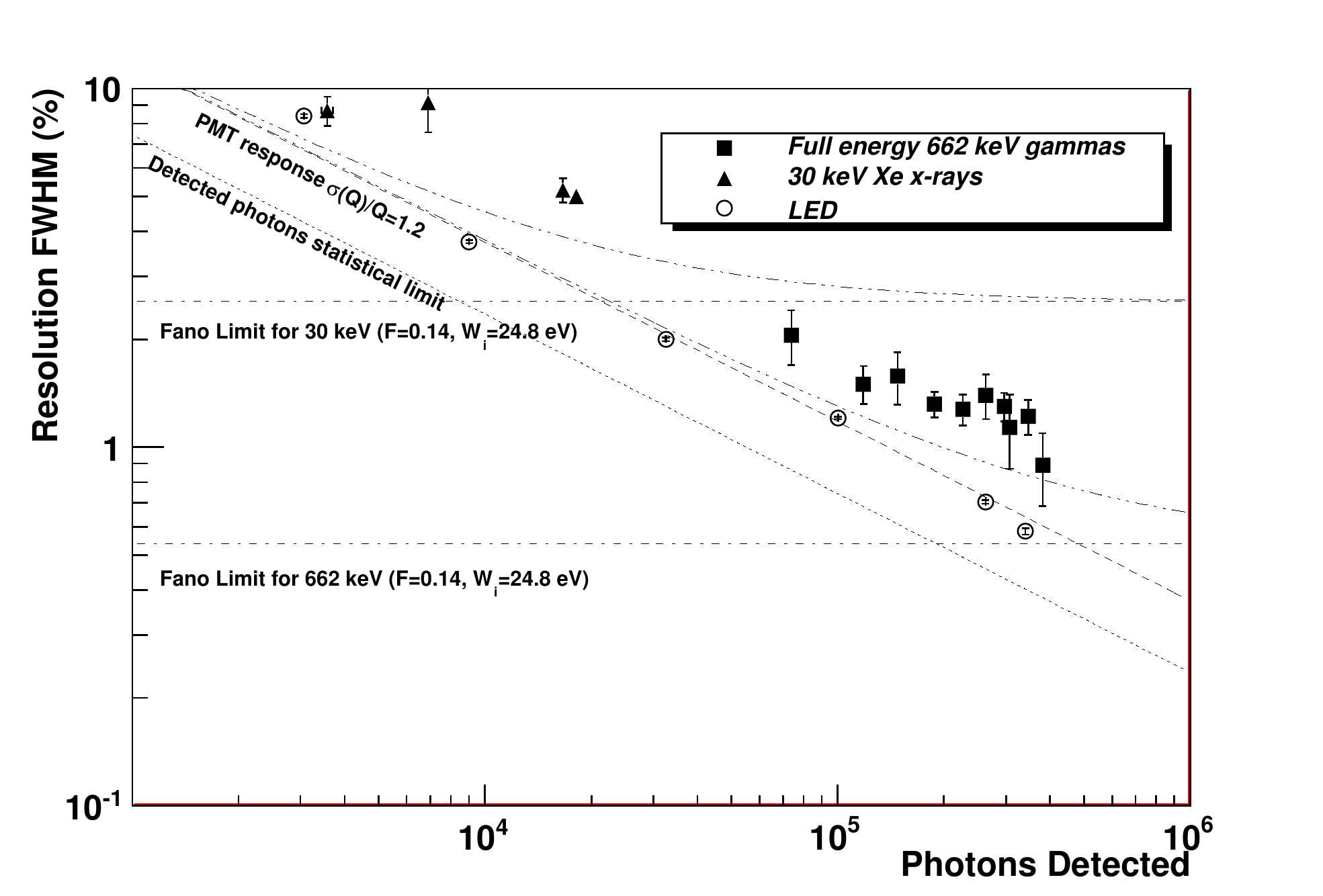}
  \caption{\label{resolution_summary} \textbf{Energy resolution in the high-pressure xenon NEXT-DBDM electroluminescent TPC:} Data points show the 
  measured energy resolution for 662 keV gammas (squares), $\sim$30 keV xenon X-rays (triangles) and LED light pulses (circles) as a function of the 
  number of photons detected. The expected resolution including the intrinsic Fano factor, the statistical fluctuations in the number of detected 
  photons and the PMT charge measurement variance is shown for  X-rays (dot dot dashed) and for 662 keV gammas (dot dot dot dashed). Resolutions 
  for the 662 keV peak were obtained from 15.1 atm data runs while X-ray resolutions we obtained from 10.1 atm runs. }
\end{figure}

Figure \ref{resolution_summary} summarizes our measurements and understanding of the EL TPC energy resolution. The lower diagonal line represents 
the Poisson statistical limit from the measurement of a small fraction of the photons produced by the EL gain while the upper diagonal line 
includes the degradation (mostly from PMT afterpulsing) due to PMT response. The circle data points show the energy resolutions obtained for 
dedicated LED runs with varying light intensities per LED pulse. The LED points follow the expected resolution over the two decades range studied. 
The two horizontal lines represent the xenon gas nominal intrinsic resolution for 30 and 662 keV, respectively, and the two curved lines are the 
expected EL TPC resolutions with contributions from the intrinsic limit and the photons' measurement. Our 662 keV data (squares) and xenon X-ray 
data (triangles) taken with various EL gains follow the expected functional form of the resolution but are 20-30\% larger possibly due to the $x-y$ 
response non-uniformity. Detailed track imaging from a dense photosensor array near the EL region, such as the one recently commissioned for 
the NEXT-DBDM prototype (see Sec. \ref{sipm_section}), will enable the application of $x-y$ position corrections to further improve the energy measurement.    
 
\begin{figure}[!htb]
  \centering
  \includegraphics[scale=0.6]{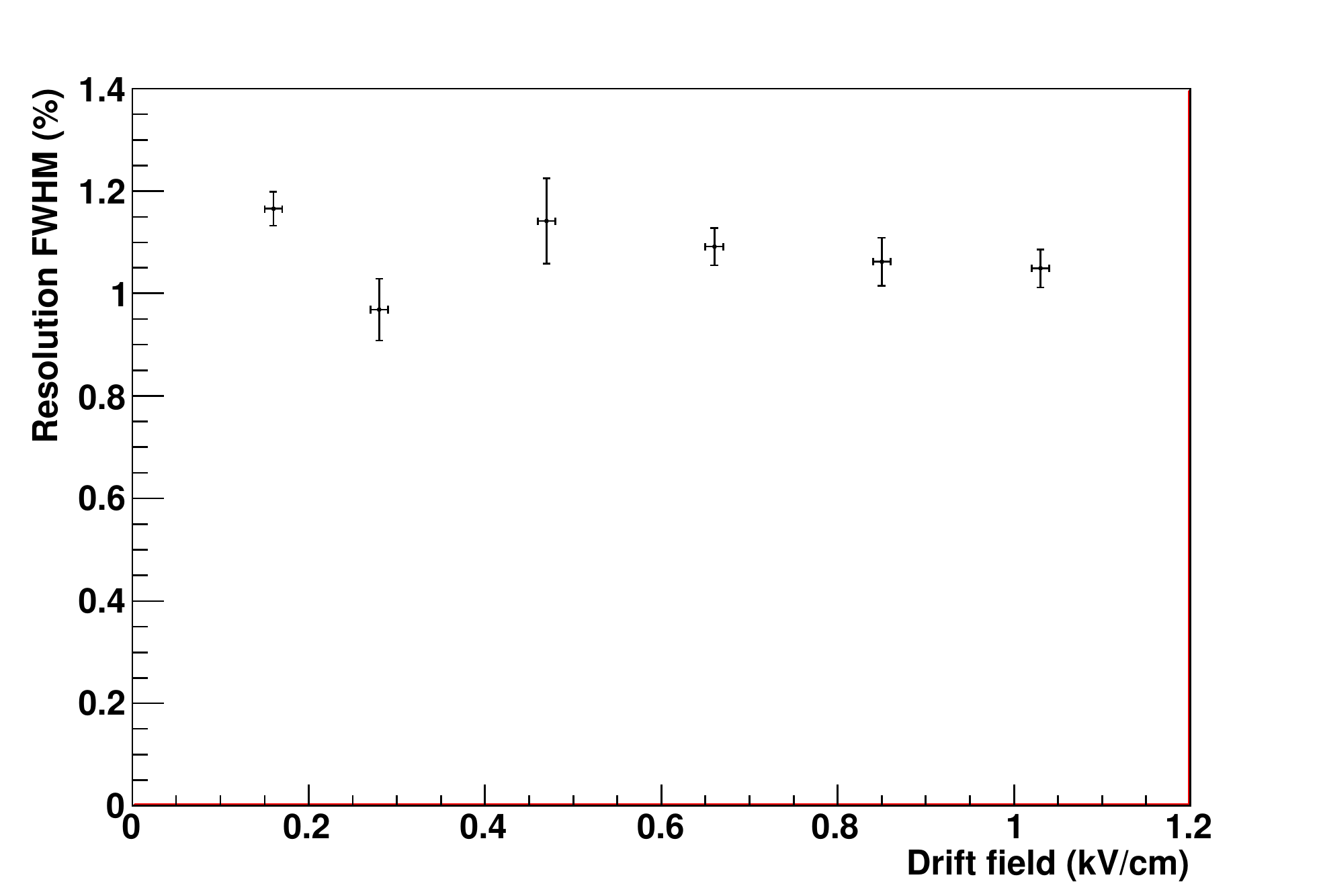}
  \caption{\label{drift_dependence_10atm_resolution} \textbf{Energy resolution dependence on drift field:} These data were taken at 10.1 atm with 
  2.1 - 2.7 kV/(cm atm) in the EL region. The energy resolution attained is largely independent of the drift electric field in the investigated 
  range. }
\end{figure}

In Ref. \cite{Bolotnikov:1997}, a study of the energy resolution for 662 keV gamma rays at pressures near condensation (30 atm and above) using a 
xenon ionization chamber found an improvement of the resolution for increased drift fields.  Asymptotic optimum resolutions were achieved only 
after applying 4 kV/cm or larger fields. Since these large fields would be difficult to achieve in a detector with one meter long drift region such 
as the one planned for NEXT-100, we investigated the resolution dependence on the drift electric field at 10 atm. Figure 
\ref{drift_dependence_10atm_resolution} shows no discernible dependence in the 0.16-1.03 kV/cm region investigated.      

\subsection{EL TPC characterization: light yield, drift velocity and longitudinal diffusion}

\begin{figure}[!htb]
  \centering
  \includegraphics[scale=0.6]{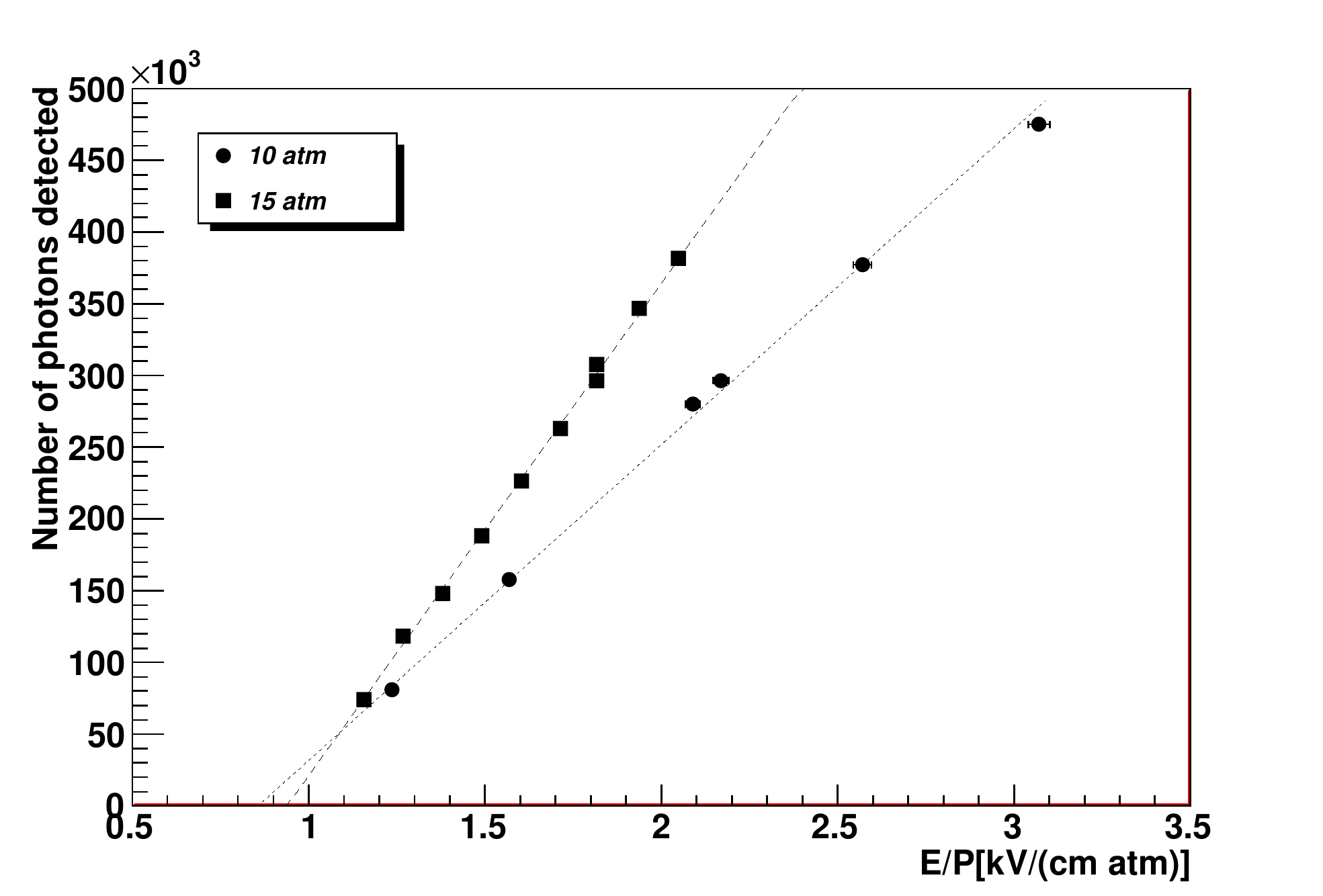}
  \caption{\label{el_dependence_yield} \textbf{Light yield dependence as a function of {\it E/P} in the EL region} }
\end{figure}

\begin{figure}[!htb]
  \centering
  \includegraphics[scale=0.6]{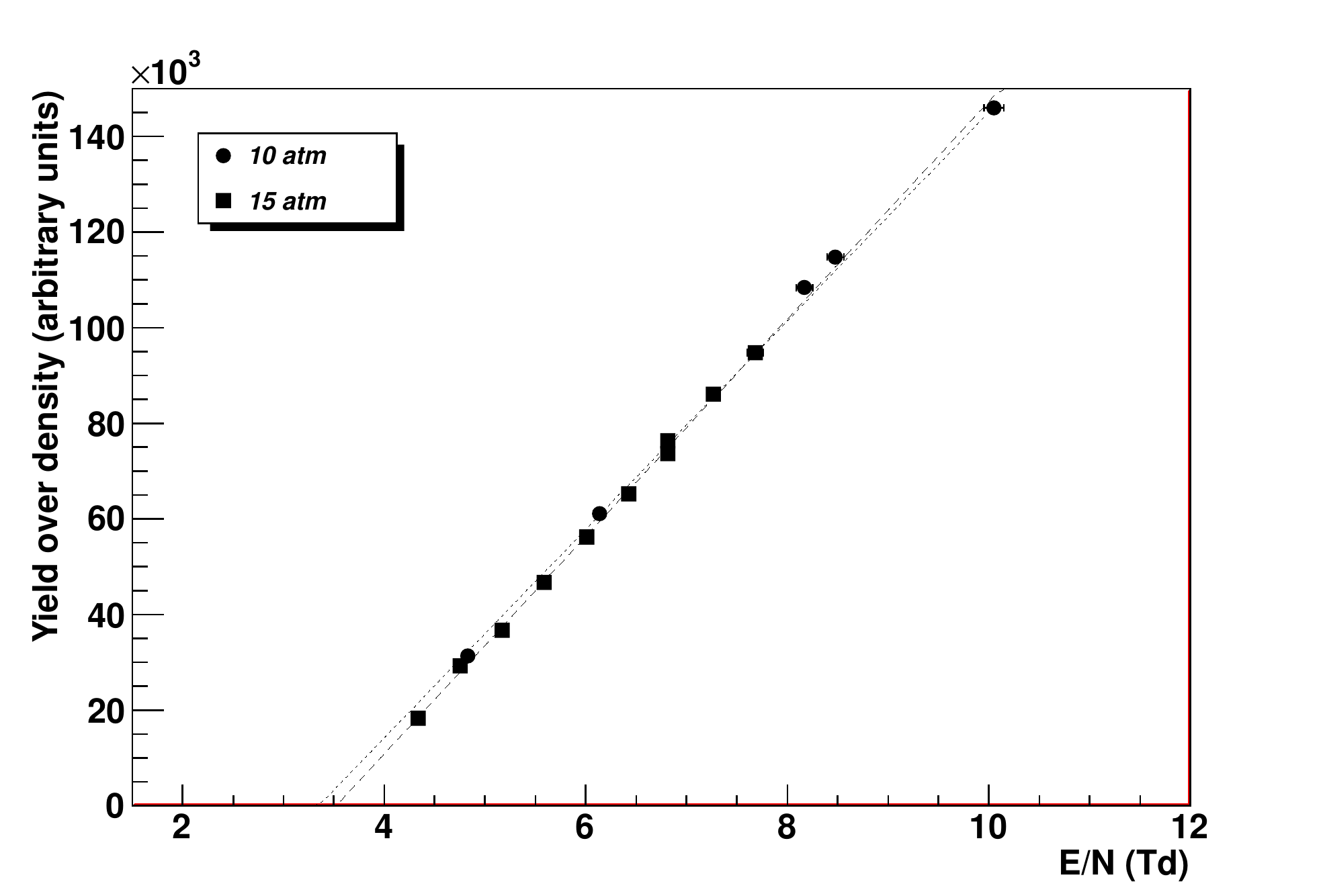}
  \caption{\label{el_dependence_yieldoverp} \textbf{Density normalized light yield vs. EL reduced field.}}
\end{figure}

While the study of the energy resolution achievable in the EL TPC was the main goal of this work, we 
pursued other measurements as well as cross checks against previous results by others. Figure \ref{el_dependence_yield}
shows the linearity of the EL light yield as the {\it E/P} in the 0.5 cm EL region is varied from 1 to 2 kV/(cm atm) at 15 atm and from 1 to 3 
kV/(cm atm) at 10 atm. At these pressures xenon deviates from the ideal gas behavior at the 10\% level. Therefore, the more appropriate 
variables to describe the process are density (N) and reduced electric field (E/N). Figure \ref{el_dependence_yieldoverp} shows the density 
normalized EL yield as a function of the reduced field. As expected, points from the 10 and 15 atm data sets follow the same linear trend with 
consistent slope and threshold.  
    
\begin{figure}[!htb]
  \centering
  \includegraphics[scale=0.6]{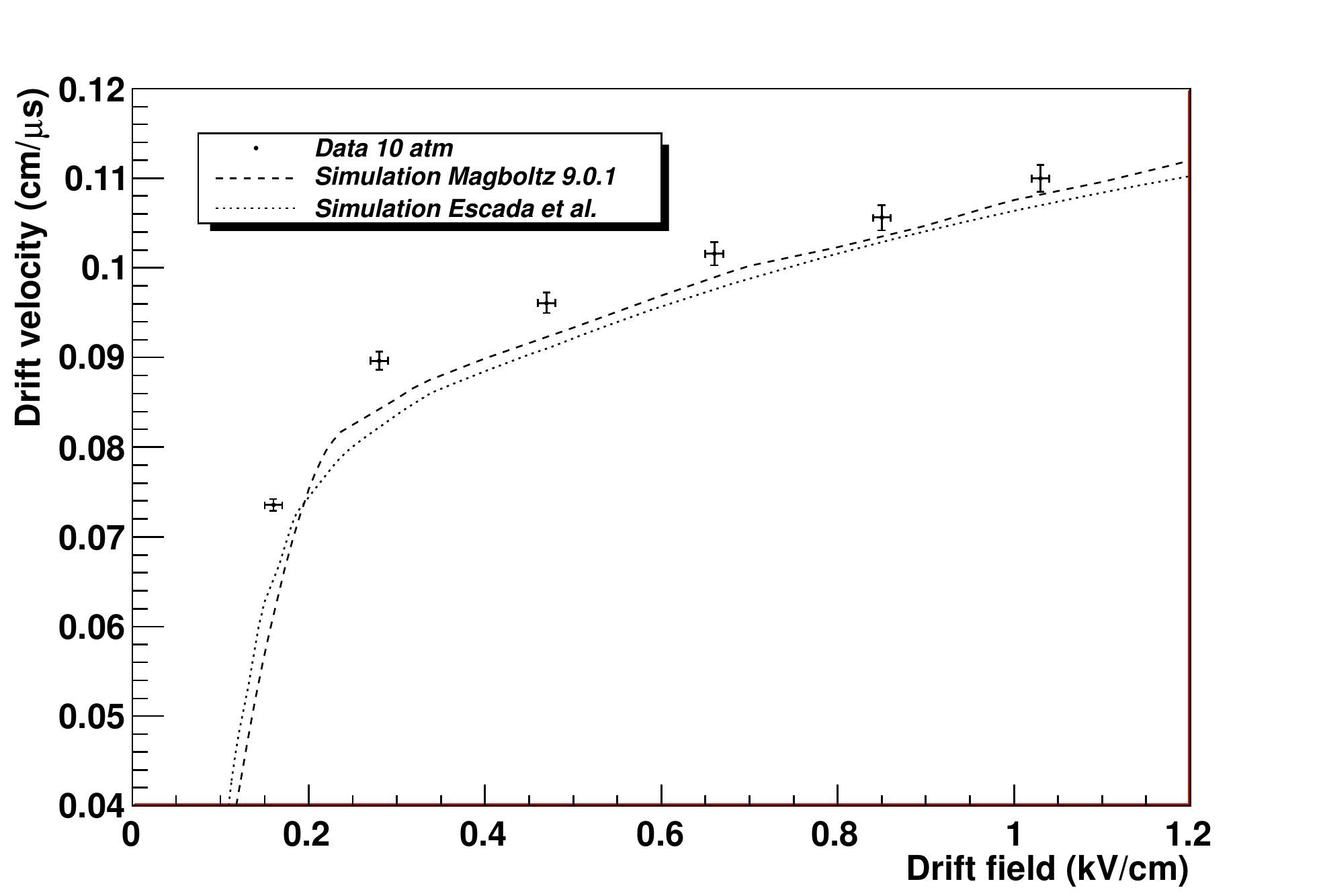}
  \caption{\label{drift_dependence_velocity} \textbf{Drift velocity dependence on drift field:} These data were taken at 10.1 atm.}
\end{figure}

The electron drift velocity was obtained from the maximum drift times measured (which correspond to the full drift length of 8 cm) for X-ray energy 
depositions. Figure \ref{drift_dependence_velocity} shows the drift velocity measured for various drift fields along with two Monte Carlo 
calculations \cite{Magboltz}\cite{Escada:2011} using up-to-date xenon-electron collision cross sections.
While the data points follow the trend in the calculations, 10\% deviations are seen at the lower drift fields. Yet, the general agreement observed 
validates the pressure and field measurements as well as the drift field uniformity and the xenon purity.

\begin{figure}[!htb]
  \centering
  \includegraphics[scale=0.6]{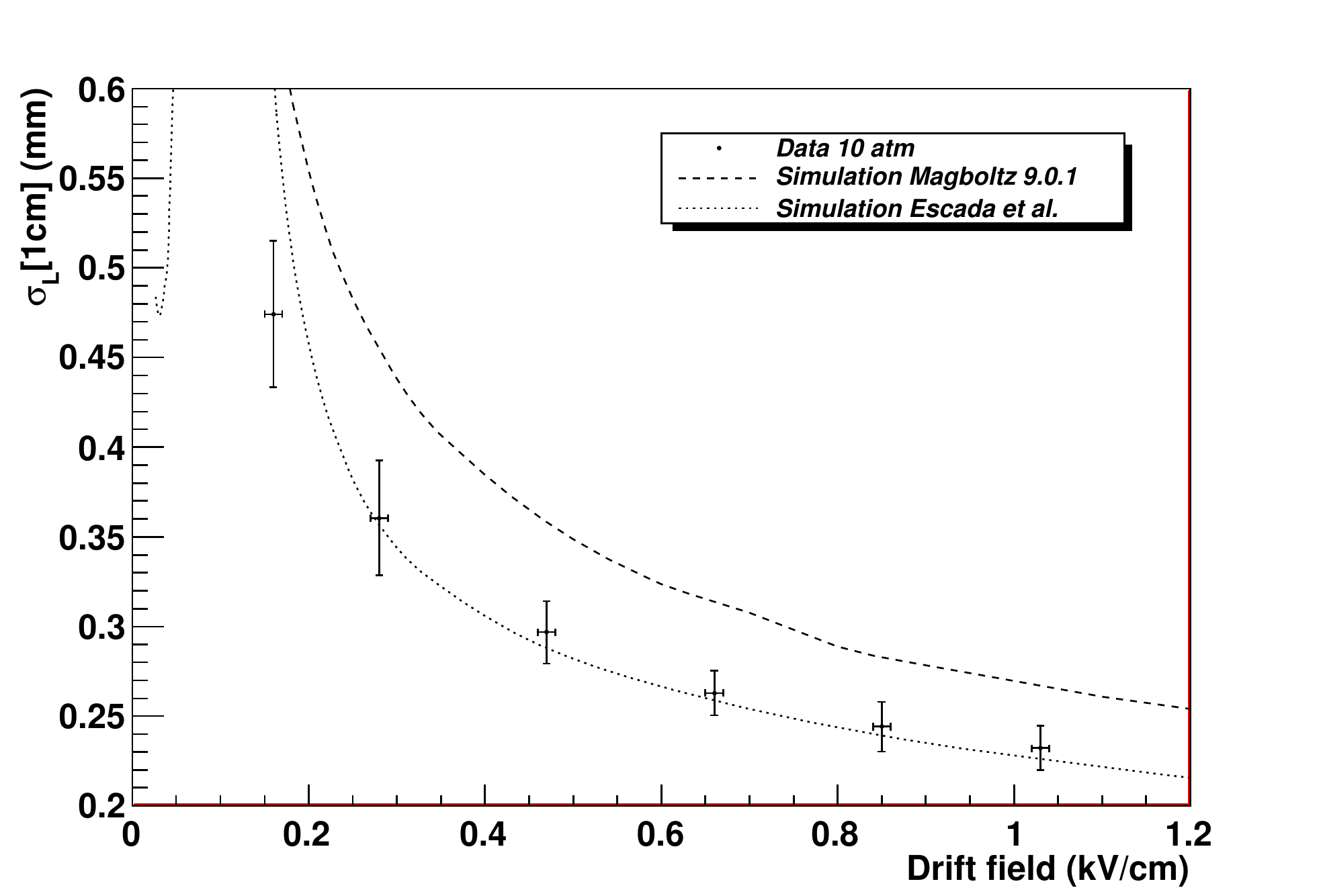}
  \caption{\label{drift_dependence_long_diffusion} \textbf{Longitudinal diffusion dependence on drift field:} The measured longitudinal diffusion 
  is presented here as the $\sigma$ spread along the drift direction for a 1 cm drift. These data were taken at 10.1 atm.}
\end{figure}

The longitudinal diffusion was calculated from the time spread of EL light of X-ray depositions (see example fits in the upper right insets of 
Figs. \ref{waveform_long_drift} and \ref{waveform_short_drift}) and using the measured drift velocity. For X-ray pulses with long drift times the 
longitudinal diffusion is the dominant source of pulse width with subdominant contributions from the transit time of electrons through the EL gap 
and from the ionization track length. Figure \ref{drift_dependence_long_diffusion} shows the dependence of the longitudinal diffusion on the drift 
field and the corresponding Monte Carlo calculations that mostly bracket the data. 
 
\clearpage 

\subsection{S1 measurement}

\begin{figure}[!htb]
\centering
\includegraphics[scale=0.6]{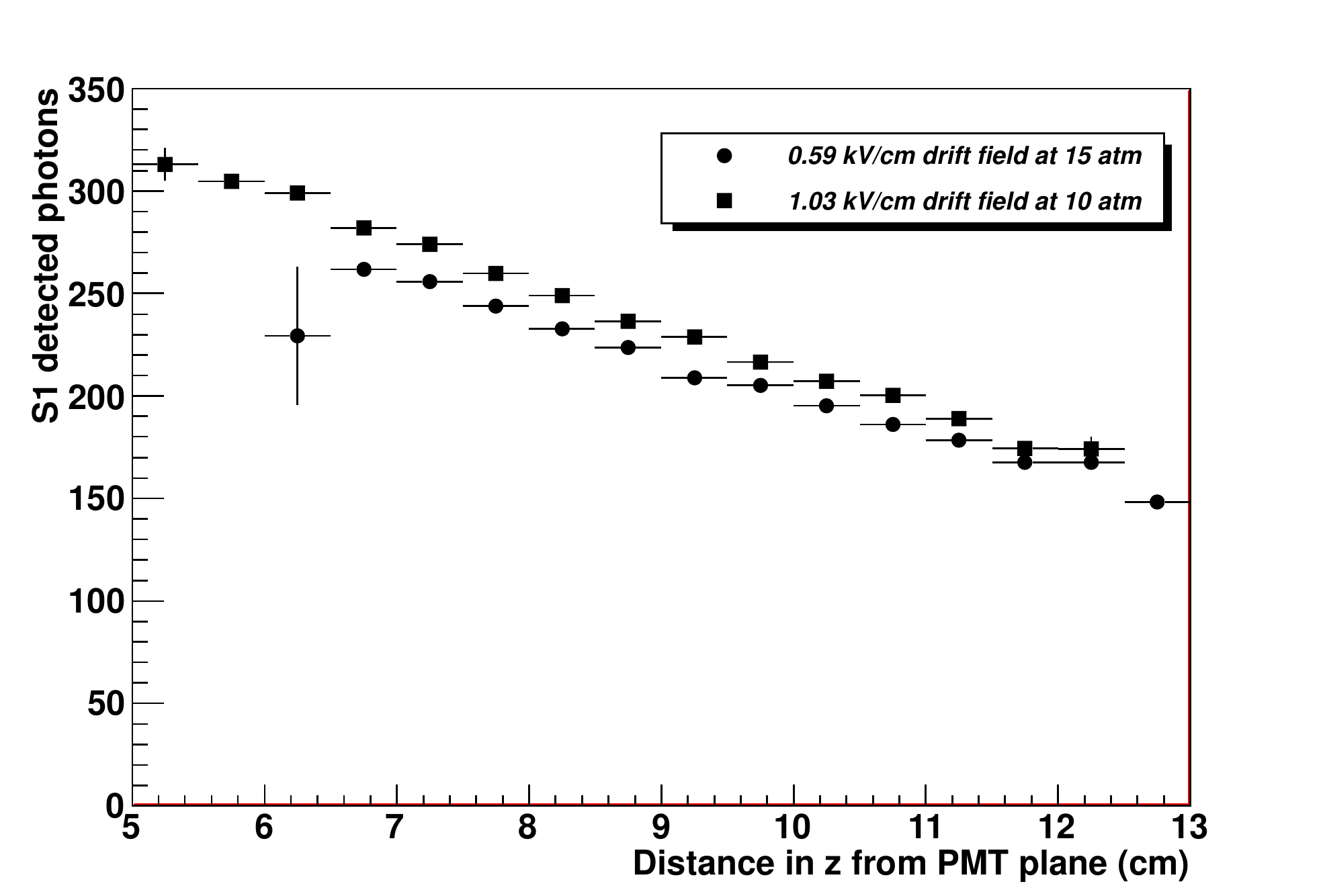}
\caption{\label{s1_position} \textbf{S1 signal dependence on z position:} For events with full energy 662 keV depositions, the number of measured S1 photons is shown versus the average distance of the ionization track from the PMT plane (obtained from the S2 charge-weighted drift time and the measured drift velocity). Two runs are shown with {\it E/P} in the \underline{drift} region of 0.04 and 0.10 kV/(cm atm) with very similar S1 yield. }
\end{figure}

Unlike the S2 light which is generated within a small 0.5 cm region in z, the S1 photons are produced throughout the entire TPC volume wherever the 
primary ionization happened. As a result the S1 charge (number of photons detected) shows large event-by-event variations. S1 light collection 
efficiency is largest for events closer to the PMT array due to the larger solid angle. Figure \ref{s1_position} shows the S1 signal's dependency 
on the average z position of the event. Between 0.2 and 0.45 S1 photons are detected per keV of energy deposited. This S1 yield is consistent with 
the known $W_\mathrm{s}$, the average energy loss required to liberate one primary scintillation photon in gaseous xenon \cite{Parsons:1990, Carmo:2008, Fernandes:2010}, 
and the photon collection and detection efficiency in our TPC.  
S1 photons are produced from direct excitations of xenon atoms by the ionizing particle and by ion-electron recombinations. The latter component 
is, in general, field and pressure dependent. Figure \ref{s1_position} shows less than 10\% differences between the S1 yield of runs taken with 
reduced fields of 0.04 and 0.10 kV/(cm atm) in the drift region. We conclude that, as expected, any recombination effects are small for electron 
energy depositions of interest here in the range of pressures and fields investigated.  

% xenon light absorption ? XXX

%% \par Effects of N2 (drift velocity, measured charge, attachment) XXX 
%% \par Head/tail measurement XXX

\section{Further Developments and Plans}
\subsection{Tracking Plane}
\label{sipm_section}
Since the completion of the energy resolution studies reported above we have
finished the construction and installation of the sensor array for the tracking plane.
It is a 1-cm-pitch 8 by 8 square pattern array of Hamamatsu S10362-11-025P silicon photomultipliers (SiPMs) 
each with 1 mm$^2$ active area. The SiPMs are coated with a thin film of tetraphenyl butadiene (TPB)
and placed in the center of the hexagonal plane 2 mm behind the electroluminescent region of the
TPC. The SiPMs produce an electronic pulse for each detected photon, and the TPB serves to shift
the incoming 172 nm light from xenon to visible light detectable by the SiPMs. Details of the SiPM performance and
coating are given in Ref. \cite{NEXT_SiPMs:2012}. The low power electronics to amplify, shape and multiplex the signals
from the array was custom built to fit inside the pressure vessel. The digitization of the multiplexed signals
is done outside the vessel with the same digitizers as for the PMT system. 

\begin{figure}[!htb]
\centering
\includegraphics[scale=0.3]{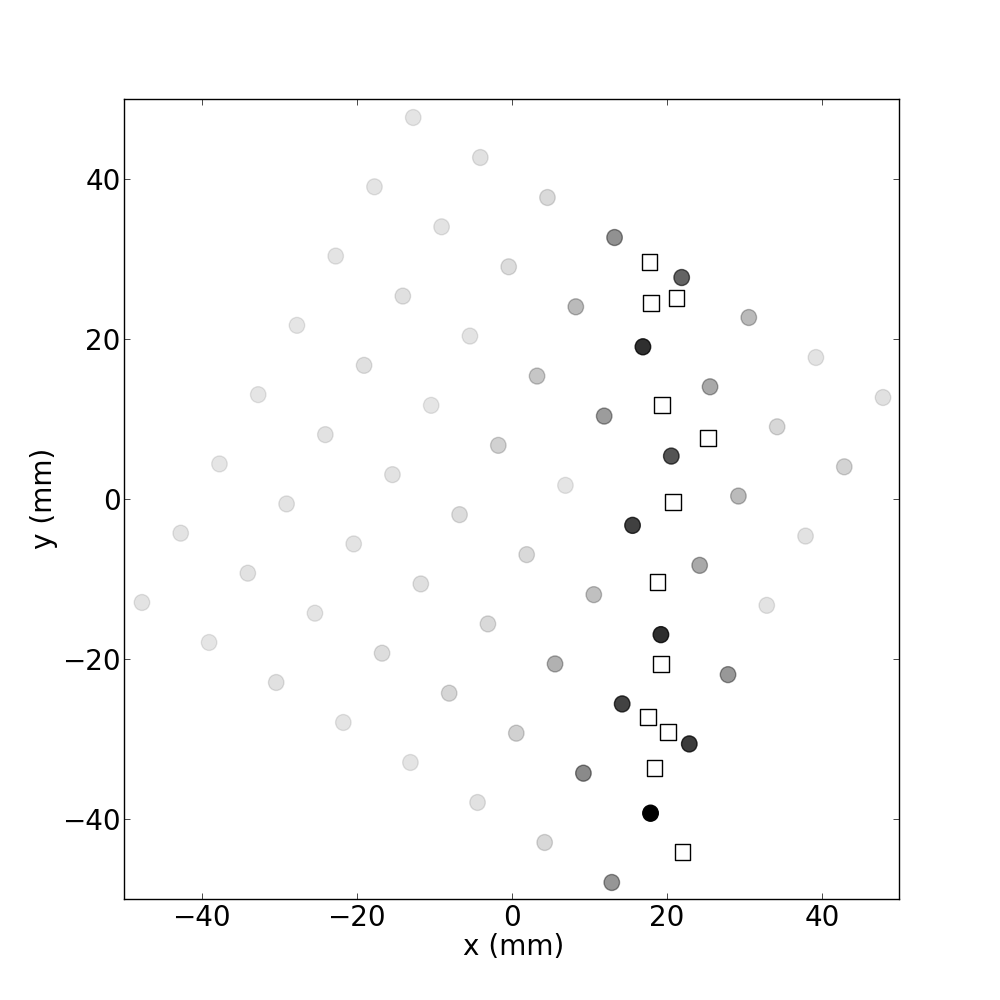}
\caption{\label{sipm_muon_array} \textbf{Cosmic ray muon track measured in the 8 x 8 SiPM array near the EL region:} 
The 64 SiPMs covering an 8 cm by 8 cm area in the center of the TPC are shown as circles shaded relatively according to the total amount of light
observed by each. The $x$-$y$ coordinates at each point in the reconstructed track are given by the squares. }
\end{figure}

\begin{figure}[!htb]
\centering
\includegraphics[scale=0.4]{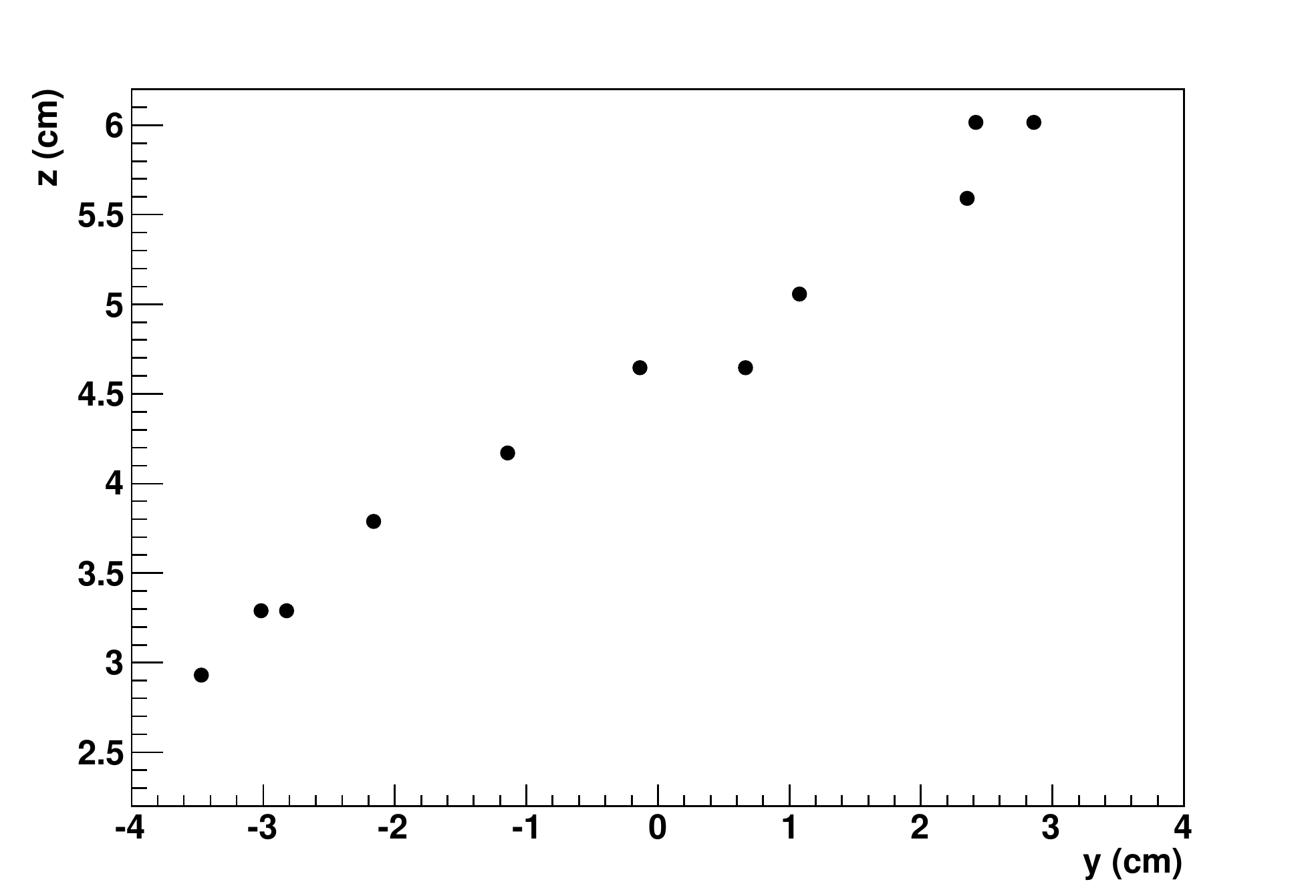}
\caption{\label{sipm_muon_2D} \textbf{Projection in $zy$ of the cosmic muon track:} The $z$ coordinate is 
calculated from the time delay between the SiPM signals and the event S1 with the known drift velocity. }
\end{figure}

Details of the operation, calibration and analysis of the SiPM array for the tracking plane will be 
reported in a future publication. Here, we present first results that demonstrate the system's ability to 
produce detailed track images.  Figure \ref{sipm_muon_array} shows the relative amounts of light detected in each SiPM
 from the electroluminescence produced by an externally tagged cosmic ray muon that traversed the 
active volume of the TPC. A reconstruction
based on a likelihood fit of the time sliced data shows the straight muon trajectory.  The RMS error in position
is estimated to be approximately 2.1 mm for individual reconstructed points.  Figure \ref{sipm_muon_2D}
shows the $z$-$y$ projection of the reconstructed points obtained using the time projection of the TPC. 
A clear 3D track is obtained even for this reconstruction that uses raw (not calibrated) SiPM signal waveforms.

\begin{figure}[!htb]
\centering
\includegraphics[scale=0.3]{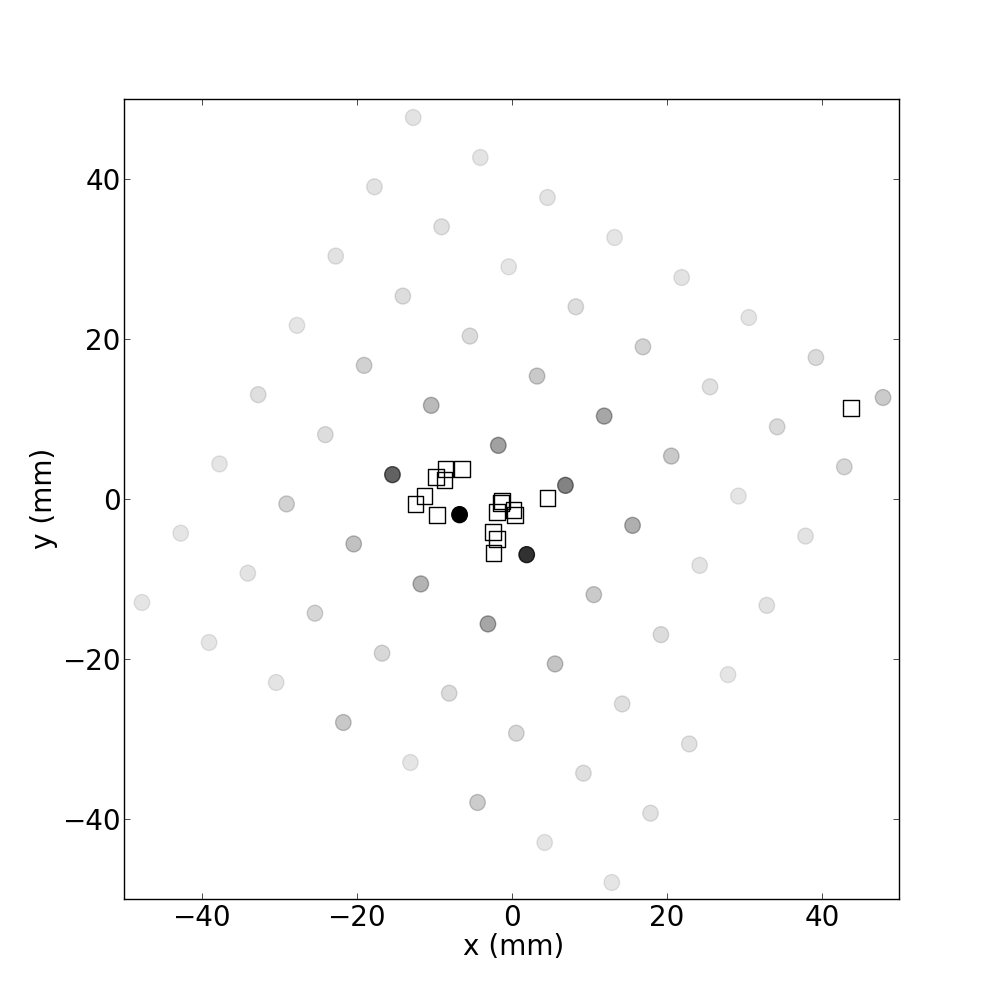}
\caption{\label{sipm_gamma_xray_2D} \textbf{Gamma ray event with photoelectron track and X-ray.}}
\end{figure}

\begin{figure}[!htb]
\centering
\includegraphics[scale=0.4]{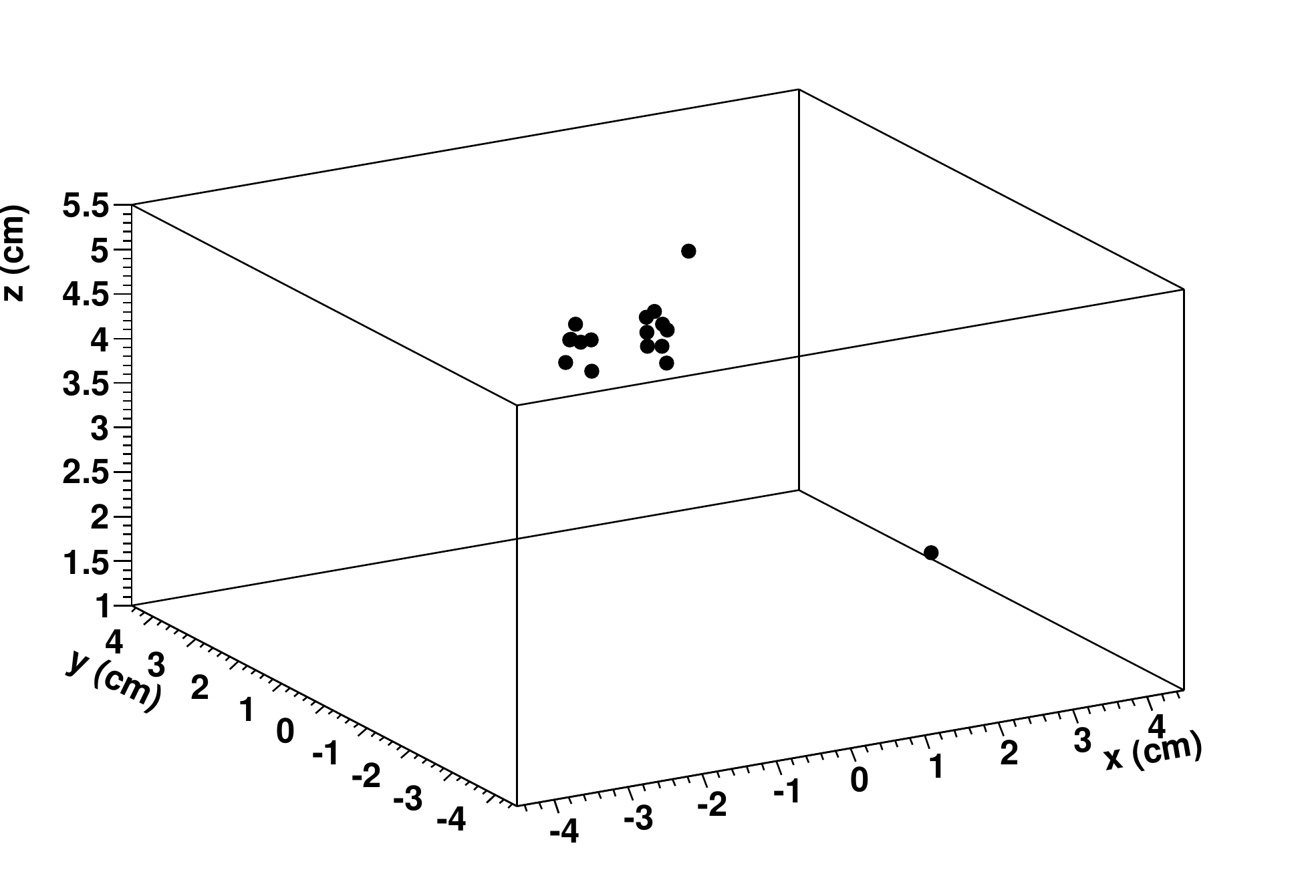}
\caption{\label{sipm_gamma_xray_3D} \textbf{3D reconstruction of gamma ray event with photoelectron track and X-ray.}}
\end{figure}

Figures \ref{sipm_gamma_xray_2D} and \ref{sipm_gamma_xray_3D} show the 2D and 3D projections 
of the track reconstruction of a full energy 662 keV gamma ray event. The 30 keV X-ray deposition is well
separated in $z$ (time) and in $x$-$y$ about 5 centimeters away from the 3 cm long photoelectron track that 
shows the expected multiple scattering.

Track information from this SiPM system will be used to map and correct the $x$-$y$ energy response 
of the PMT plane with the goal of demonstrating a near-intrinsic energy resolution over a large part of the 
active volume of the detector. SiPM tracks will also be used to study the topological information that will 
enable, in the NEXT-100 detector, the identification of $0\nu\beta\beta$ candidate events while rejecting background single and multi-track events from 
gamma rays from residual radioactivity in detector and surrounding materials.

\subsection{Gas Additives and Nuclear Recoil Studies}

A research program is underway to study gas additives to optimize the detector performance.
We are investigating the properties of high pressure xenon mixtures with CF$_4$, CH$_4$ and TMA (trimethylamine).
These studies focus on diffusion (for track imaging), electroluminescence and primary scintillation yield, and energy resolution and will be reported elsewhere.

We are also investigating the high pressure EL TPC response to low energy nuclear recoils. O(10 keV) nuclear 
recoils are expected from WIMP (Weakly Interacting Massive Particle) dark matter elastic interactions with the
xenon target.  These studies, using MeV neutron sources, aim to characterize the S2/S1 discrimination 
between electron and nuclear recoils and to optimize the light collection from the feeble nuclear recoil S1 signals.  

\section{Conclusions}
We presented the design, data and results from the NEXT-DBDM high-pressure gaseous natural xenon electroluminescent TPC 
that was built at the Lawrence Berkeley National Laboratory. Energy resolutions of 1\% FWHM for 662 keV gamma 
rays were obtained at 10 and 15 atm and 5\% FWHM for 30 keV fluorescence xenon X-rays. The main contributions to the 
energy resolution were studied and are well understood and thus justify the extrapolation to a 0.5\% FWHM resolution for the 2,459 keV hypothetical neutrino-less double beta decay peak. This energy resolution is a factor 7 to 20 better than that of the current leading $0\nu\beta\beta$ experiments using liquid xenon. We presented also first results from a track imaging system consisting of 64 silicon photo-multipliers recently installed in NEXT-DBDM that, along with the excellent energy resoution, demonstrates the key functionalities required for the NEXT-100 $0\nu\beta\beta$ search. The construction of the NEXT-100 detector with 90 kg of $^{136}$Xe detector has began and first data is expected for 2015.   

\section{Acknowledgements}
% NS-DOE
We thank Adam Bernstein and Mike Heffner for the loan of the high pressure and storage vessels from LLNL
(under LLNL loan 101-3026). This work was supported by the Director, Office
 of Science, Office of Basic Energy Sciences, of the U.S. Department of Energy under Contract No. DE-AC02-05CH11231.
This work used resources of the National Energy Research Scientific Computing Center (NERSC).
J.~Renner (LBNL) acknowledges the support of a US DOE NNSA Stewardship Science Graduate Fellowship under
contract no. DE-FC52-08NA28752.

%% The Appendices part is started with the command \appendix;
%% appendix sections are then done as normal sections
%% \appendix

%% \section{}
%% \label{}

%% References
%%
%% Following citation commands can be used in the body text:
%% Usage of \cite is as follows:
%%   \cite{key}         ==>>  [#]
%%   \cite[chap. 2]{key} ==>> [#, chap. 2]
%%

%% References with bibTeX database:
\section*{References}
\bibliographystyle{model1-num-names}
\bibliography{xenon_tpc}

\end{document}